\documentclass[]{aa}
\usepackage{natbib,graphics,graphicx,psfig,amssymb,rotfloat,amssymb,amsmath}
\bibpunct{(}{)}{;}{a}{}{,}

\newcommand{\lsun}{\ensuremath{\mathrm{L}_{\odot}}}

\newcommand{\msun}{\ensuremath{\mathrm{M}_{\odot}}}
\newcommand{\msunyr}{\ensuremath{\mathrm{M}_{\odot} {\rm yr}^{-1}}}

\begin{document}

\title{A 10 $\mu$m spectroscopic survey of Herbig Ae star disks:
grain growth and crystallization\thanks{Based on
        observations obtained at the European Southern
        Observatory (ESO), La Silla, observing programmes
        68.C-0536 and 70.C-0533,
        and on observations with ISO, an ESA project with instruments
        funded by ESA Member States (especially the PI countries: France,
        Germany, the Netherlands and the United Kingdom) and with the
        participation of ISAS and NASA.}}

\author{
R.~van~Boekel\inst{1,2}
\and M.~Min\inst{1}
\and L.B.F.M.~Waters\inst{1,3}
\and A.~de Koter\inst{1}
\and C.~Dominik\inst{1}
\and M.E.~van~den~Ancker\inst{2}
\and J.~Bouwman\inst{4}
}
\institute{
Astronomical Institute ``Anton Pannekoek'', University of
Amsterdam, Kruislaan 403, NL--1098 SJ  Amsterdam, The Netherlands
\and
European Southern Observatory, Karl-Schwarzschildstrasse 2, D-85748
Garching bei M\"unchen, Germany
 \and
Instituut voor Sterrenkunde, Katholieke Universiteit Leuven,
Celestijnenlaan 200B, B-3001 Heverlee, Belgium
 \and
Max-Planck-Institut f\"ur Astronomie, K\"onigstuhl 17,
69117 Heidelberg, Germany
}

\offprints{R. van Boekel (vboekel@science.uva.nl)}
\date{Received 2004 November 9; accepted 2005 March 4}

\abstract{
We present spectroscopic observations of a large sample of Herbig Ae
stars in the 10\,$\mu$m spectral region. We perform compositional fits
of the spectra based on properties of homogeneous as well as
inhomogeneous spherical particles,
and derive the mineralogy and typical grain sizes of the
dust responsible for the 10\,$\mu$m emission. Several trends are
reported that can constrain theoretical models of dust processing in
these systems: {\it i}) none of the sources consists of fully pristine dust
comparable to that found in the interstellar medium, {\it ii}) all sources
with a high fraction of crystalline silicates are dominated by large
grains, {\it iii}) the disks around more massive stars 
(M\,$\gtrsim$\,2.5\,\msun, L\,$\gtrsim$\,60\,\lsun)
have a higher fraction of crystalline silicates than those around
lower mass stars, {\it iv}) in the subset of lower mass stars
(M\,$\lesssim$\,2.5\,\msun) there is no correlation between stellar
parameters and the derived crystallinity of the dust. The
correlation between the shape and strength of the 10 micron
silicate feature reported
by \cite{2003A&A...400L..21V} is reconfirmed with
this larger sample. The evidence presented in this paper is combined
with that of other studies to present a likely scenario of dust
processing in Herbig Ae systems. We conclude that the present
data favour a scenario in which the crystalline silicates
are produced in the innermost regions of the disk, close to the star,
and transported outward to the regions where they can be detected by
means of 10 micron spectroscopy. Additionally, we conclude that the final
crystallinity of these disks is reached very soon after active accretion
has stopped.}

\authorrunning{Van Boekel et al.}
\titlerunning{Grain processing in Herbig Ae/Be stars}

\maketitle


\section{Introduction}
\label{sec:t2_fs_introduction}
A characteristic shared by many young, low and intermediate mass stars
is the presence of a strong infrared excess. This radiation emerges
from circumstellar dust grains left over from the star formation
process. The material is believed to reside in a disk which is formed
as a result of angular momentum conservation in the collapsing
molecular cloud. After an initial strong accretion phase, a much longer
pre-main-sequence phase ensues during which the disk slowly
dissipates, and possibly planets are formed.

The interstellar dust which finds its way into a proto-planetary disk
will undergo large changes in average size and chemical composition.
These changes trace the process of disk dissipation and planet
formation. Our solar system contains a precious record of the processes
that took place during its formation. Comparison of this record to what
is observed around young pre-main-sequence stars provides important
insight into the history of our own solar system, and helps to
constrain planet formation models.

The infrared spectral region is rich in vibrational resonances of
abundant dust species. Therefore, infrared spectroscopy can be used to
determine the composition of dust in proto-planetary disks,
as well as constrain the size and shape of the dust grains. 
Analysis of the infrared (IR) dust emission features originating from the
disk surface layer can
be used
to establish to what extent the dust composition in the disk has
evolved away from that seen in the interstellar medium (ISM). For
instance, crystalline silicates appear absent in the ISM
\citep[e.g.][]{2000ESASP.456..183D,2004ApJ...609..826K}
but are a substantial component in (some) comets and in
interplanetary dust particles
\citep{1987RvGeo..25.1527M,1992ApJ...394..643B}. Clearly, the
refractory material in the proto-solar cloud went through large changes
as the solar system was formed.
It should be kept in mind that spectroscopy in the 10~micron
region is sensitive to a
limited grain size range: large grains (with sizes above a few $\mu$m,
depending on chemical composition and wavelength) show only weak
spectral structure and do not contribute significantly to the
infrared emission features.

In this work we study the composition of dust in the circumstellar
environment of Herbig Ae/Be stars
\citep[][]{1960ApJS....4..337H},
using infrared spectroscopy. We restrict our study to
a sub-group of mostly late B and A-F type stars (hereafter
HAe stars). These stars show little or no optical extinction and low mass
accretion rates, as derived from radio analysis 
\citep{1993ApJS...87..217S}, and the lack of significant veiling
in optical spectra. For these lower mass members of the Herbig class,
evidence for the disk hypothesis is compelling 
\citep[e.g.][]{1997ApJ...490..792M,2001AJ....122.3396G,2001A&A...365...78A,
2003ApJ...588..360E}.

The observed SEDs of these HAe stars can very well be explained with
models for gas-rich, passively heated disks in hydrostatic equilibrium
and a puffed-up inner rim \citep{2001ApJ...560..957D}. These models
indicate that the emission observed at near-IR wavelengths is
dominated by the inner rim, while the mid-IR spectrum has a large
contribution from the dust grains in the warm surface layers of the
disk, typically located between a few to several tens of Astronomical
Units (AU) from the star. Therefore, mid-IR spectroscopy predominantly
relates to the composition of the dust in this surface layer. 
Van Boekel et al. (\citeyear{2003A&A...400L..21V})
argue that due to the turbulent nature of the disks, the small grains
observed at the disk surface are well coupled to those in the disk
mid-plane, and so the observations of the surface layers likely bear
relevance for the overall small grain population.

Observations of the dust composition in HAe disks have revealed a very
rich mineralogy, and strong source to source variations of the dust
composition \citep[e.g.][]{2001A&A...375..950B}.
Some stars show strong 9.7 and 18\,$\mu$m amorphous
silicate emission, with a band shape and strength very similar to that
seen in the interstellar medium. Other objects have only weak silicate
emission, with sub-structure at 9.2, 10.6 and 11.3\,$\mu$m due to
crystalline silicates. At longer wavelengths, high-quality observations
only exist for a small number of stars, so far all obtained with the Infrared Space 
Observatory (ISO). This situation is rapidly improving due to 
ongoing observations with the Infrared Spectrograph
on board of the Spitzer Space Telescope. 
The ISO spectra reveal a
mineralogy dominated by the crystalline silicates forsterite and
enstatite, i.e. Mg-rich, Fe-poor materials 
\citep[e.g.][ hereafter ME01]{1998A&A...332L..25M,{2001A&A...375..950B},
2001A&A...365..476M}. A small group of
stars lacks silicate emission, but shows prominent emission from
Polycyclic Aromatic Hydrocarbons (PAHs) at 
3.3, 6.2, 7.7-7.9, 8.6 and 11.3\,$\mu$m. Many stars show a combination
of silicates and PAHs.

In recent years, several investigators have attempted to find
correlations between the properties of the dust in the disk of Herbig
stars on the one hand, and global properties of the disk (e.g. disk
geometry) and/or the star (e.g. mass, luminosity, age, binarity) on
the other hand \citep[ME01;][]{2001A&A...375..950B, 2004A&A...426..151A}.  
Perhaps the most
promising results so far are relations between the dust properties and
the shape of the SEDs of the disk; for instance, the PAH bands are on
average stronger in sources with SEDs that rise at far-IR wavelengths
(ME01).
This has been interpreted in terms of the disk geometry in the
following way. Relatively red SEDs correspond to flaring disks, that have a large
surface which is directly irradiated by the star. If PAHs are present
in this flaring disk surface layer, they will contribute to the
emission in the familiar PAH bands.
Van Boekel et al. (\citeyear{2004A&A...418..177V}) studied the spatial
distribution of the PAH emission in HD\,97048, which was found to be
extended on a scale of 1-2\,arcsec ($\sim$250\,AU)
but clearly associated with the
disk. It is likely that also in the other stars of our sample, the PAH
emission originates from the outer disk region.
Disks that lack a flaring outer
region will have much less prominent PAH emission. \cite{2004A&A...426..151A}
confirm this relation between PAH band strength and SED shape using a
sample of about 50 Herbig Ae/Be stars.  \cite{2004A&A...422..621A} note that
stars with flat far-IR SEDs on average have flat millimeter spectral
slopes, suggesting that the cold mid-plane grains in these sources
have grown to larger size than the corresponding grains of stars with
rising far-IR SEDs.

Despite considerable efforts, it has proven difficult so far to
determine relations between stellar and dust parameters. This has
prompted us to carry out a comprehensive spectral survey at 10\,$\mu$m
of isolated HAe stars. 
Our goal is to establish relations between star and dust
properties that are of relevance for our understanding of the
evolution of dusty disks around young stars, by
increasing the number of stars for which mid-IR spectra are available.
We have used the
\emph{Thermal Infrared Multi Mode Instrument~2}
\citep[TIMMI2,][]{1998SPIE.3354..865R}, attached to the 3.6\,m
telescope of the European Southern Observatory for our spectral
survey.  All known optically bright HAe stars accessible from the La
Silla observatory were included in our initial list of targets. In
practice, high quality data could be obtained for stars with
10\,$\mu$m fluxes of about 3\,Jy or more.

Here we present the results of our spectroscopic survey at 10\,$\mu$m.
In total, we obtained high quality spectra of a sample of 24 HAe
stars, introduced in section~\ref{sec:t2_fs:sample}.  We report on the
observations and data reduction in
section~\ref{sec:t2_fs:observations_data_reduction}.  An overview of
the spectra is given in section~\ref{sec:t2_fs:description}. In
section~\ref{sec:t2_fs:analysis} we present compositional fits to the
silicate feature observed in most of the sources, using the optical
properties of minerals commonly found in circumstellar disks.  In
section~\ref{sec:t2_fs:discussion} we discuss the implications of our
results for our understanding of dust processing in HAe disks.

\section{The sample stars}
\label{sec:t2_fs:sample}

\begin{table*}
\begin{center}

\caption{Basic parameters of the sample stars. In
columns\,1 and~2 we give the index number by which a star can be
identified in tables and figures throughout this work, and the name of
the star, respectively.  The classification of the sources according
to ME01 is listed in the third column.  In column\,5 the distances,
derived from direct (Hipparcos) parallax measurements or by
association to a star forming region (SFR, column\,4), are given.
For the stars where the distance is determined by association with
a SFR we assume an error of 30\,\% in the distance.
The
spectral type according to the MK classification, the effective
temperature and the stellar luminosity are given in columns\,6, 7
and\,8, respectively.  Mass and age estimates for most stars, as
derived by comparing their positions in the HR diagram to theoretical
pre-main-sequence tracks, are given in columns\,9 and 10.
For HD\,101412 we have no reliable distance estimate, and can therefore
not determine its luminosity, mass and age.}
\label{tab:t2_fs:source_list}

\linespread{1.1}
\selectfont

\begin{tabular}{rlrlllcccc}

\hline
\vspace{-0.3cm}
\\
\vspace{0.1cm}
(1) & \ \ \ \ (2) & \ \ (3) & \ (4) & \ (5) & \ \ (6) \ & \ \ (7) & \ (8) & \ (9)  & (10) \\
\#  & \ \ \ \ star&group   &  SFR       &  d    &  Sp.Type  &  log $\mathrm{T_{eff}}$ & log L       & Mass       &  log(Age)     \\ 
  &             &   &            & [pc]  &           &      [K]   &     [\lsun] & [\msun]     &      [yr]     \\ 
\hline
\vspace{-0.2cm}
\\

  1 &    AB Aur       & Ia  & L519       &$  144_{ -   17}^{ +   22} $ & A0Ve+sh     & 3.979 &   1.67   &     2.4$\pm$0.2  &  6.3$\pm$0.2  \\ 
  2 &    UX Ori       &IIa  & Orion OB1a &$  340\pm  102$              & A4IVe       & 3.925 &   1.68   &     2.5$\pm$0.3  &  6.3$\pm$0.4  \\ 
  3 & HD\,36112       & Ia  &            &$  204_{ -   39}^{ +   63} $ & A5IVe       & 3.911 &   1.35   &     2.0$\pm$0.3  &  6.5$\pm$0.3  \\ 
  4 &    HK ORI       &IIa  & Orion OB1a &$  340\pm  102$              & A4pevar     & 3.927 &   0.87   &     1.7$\pm$0.3  & $>$6.8        \\ 
  5 & HD\,245185      & Ia  & Orion OB1a &$  340\pm  102$              & A0Ve        & 3.979 &   1.26   &     2.2$\pm$0.3  & $>$6.8        \\ 
  6 &    V380 Ori     &IIa  & Orion OB1c &$  510\pm  153$              & A1:e        & 3.965 &   1.88   &     2.8$\pm$0.5  &  6.2$\pm$0.4  \\ 
  7 & HD\,37357       &IIa  &            &$  240\pm   72^1$            & A2Ve        & 4.021 &   1.47   &     2.4$\pm$0.4  & $>$7.0        \\ 
  8 & HD\,37806       &IIa  & Orion OB1b &$  470\pm  141$              & A2Vpe       & 3.953 &   2.13   &     3.6$\pm$0.8  &  5.9$\pm$0.4  \\ 
  9 & HD\,95881       &IIa  & Sco OB2-4? &$  118\pm   35$              & A2III/IVe   & 3.954 &   0.88   &     1.7$\pm$0.2  & $>$6.5        \\ 
  10& HD\,97048       & Ib  & Ced 111    &$  175_{ -   20}^{ +   26} $ & B9.5Ve+sh   & 4.000 &   1.64   &     2.5$\pm$0.2  & $>$6.3        \\ 
  11& HD\,100453      & Ib  &            &$  111_{ -    8}^{ +   10} $ & A9Ve        & 3.869 &   0.90   &     1.7$\pm$0.2  &  7.0$\pm$0.1  \\ 
  12& HD\,100546      & Ia  & Sco OB2-4? &$  103_{ -    6}^{ +    6} $ & B9Vne       & 4.021 &   1.51   &     2.4$\pm$0.1  & $>$7.0        \\ 
  13& HD\,101412      &IIa  &            &                             & B9.5V       &       &          &                  &               \\ 
  14& HD\,104237      &IIa  & Cha III    &$  116_{ -    6}^{ +    7} $ & A4IVe+sh    & 3.925 &   1.54   &     2.3$\pm$0.1  &  6.3$\pm$0.1  \\ 
  15& HD\,135344      & Ib  & Sco OB2-3  &$  140\pm   42$              & F4Ve        & 3.819 &   0.91   &     1.6$\pm$0.2  &  6.9$\pm$0.3  \\ 
  16& HD\,139614      & Ia  & Sco OB2-3  &$  140\pm   42$              & A7Ve        & 3.895 &   0.91   &     1.7$\pm$0.3  & $>$7.0        \\ 
  17& HD\,142666      &IIa  & Sco OB2-2  &$  145\pm   43$              & A8Ve        & 3.880 &   1.13   &     1.8$\pm$0.3  &  6.8$\pm$0.4  \\ 
  18& HD\,142527      & Ia  &            &$  198_{ -   37}^{ +   60} $ & F7IIIe      & 3.796 &   1.46   &     2.5$\pm$0.3  &  6.0$\pm$0.4  \\ 
  19& HD\,144432      &IIa  & Sco OB2-2  &$  145\pm   43$              & A9IVev      & 3.866 &   1.01   &     1.8$\pm$0.2  &  7.0$\pm$0.3  \\ 
  20& HD\,144668      &IIa  & Lupus 3    &$  207_{ -   31}^{ +   45} $ & A5-7III/IVe & 3.899 &   1.94   &     3.2$\pm$0.5  &  5.7$\pm$0.3  \\ 
  21& HD\,150193      &IIa  & Sco OB2-2  &$  150_{ -   30}^{ +   50} $ & A2IVe       & 3.953 &   1.38   &     2.3$\pm$0.2  & $>$6.3        \\ 
  22& HD\,163296      &IIa  &            &$  122_{ -   13}^{ +   16} $ & A3Ve        & 3.941 &   1.38   &     2.0$\pm$0.2  &  6.7$\pm$0.4  \\ 
  23& HD\,169142      & Ib  & Sco OB2-1  &$  145\pm   43$              & A5Ve        & 3.914 &   1.16   &     2.0$\pm$0.3  &  6.9$\pm$0.3  \\ 
  24& HD\,179218      & Ia  & (L693)     &$  243_{ -   43}^{ +   68} $ & B9e         & 4.021 &   2.00   &     2.9$\pm$0.5  &  6.1$\pm$0.4  \\ 

\hline
\vspace{-0.3cm}
\end{tabular} \\
\linespread{1.0 }%
\selectfont

\footnotesize{$^1$ We adopted a distance of 240 pc so that the luminosity of this star matches its spectral type.}

\end{center}
\end{table*}

\begin{table*}
\begin{center}

\caption{
Literature infrared photometry used in this work. 
Columns\,3-7 list the magnitudes in the J~(1.25\,$\mu$m),
H~(1.65\,$\mu$m), K~(2.2\,$\mu$m), L~(3.6\,$\mu$m), and M~(4.8\,$\mu$m)
photometric bands, with references in column~8. The listed reference 
codes are:
 BO: Bouchet et al. \citeyear{1991A&AS...91..409B};
 CA: Carter \citeyear{1990MNRAS.242....1C};
 CO: Cohen \citeyear{1973MNRAS.161...97C};
 CU: Cutri et al. \citeyear{2003yCat.2246....0C};
 DW: de Winter et al. \citeyear{2001A&A...380..609D};
DW2: de Winter et al. \citeyear{1996A&AS..119....1D};
 EI: Eiroa et al. \citeyear{2001A&A...365..110E};
 FO: Fouque et al. \citeyear{1992A&AS...93..151F};
 GL: Glass \& Penston \citeyear{1974MNRAS.167..237G};
 HI: Hillenbrand et al. \citeyear{1992ApJ...397..613H};
 LA: Lawrence et al. \citeyear{1990AJ.....99.1232L};
 MA: Malfait et al. \citeyear{1998A&A...331..211M};
 ME: Mendoza \citeyear{1967BOTT....4..149M};
 ST: Strom et al. \citeyear{1990ApJ...362..168S};
 SY: Sylverster et al. \citeyear{1996MNRAS.279..915S};
 VR: Vrba et al. \citeyear{1976AJ.....81..317V};
 WA: Waters et al. \citeyear{1988A&A...203..348W}.
Columns\,9-12 contain the infrared fluxes in Jy from the IRAS Point Source
Catalogue \citep{1988IRASP.C......0J}.
}
\label{tab:t2_fs:literature_photometry}

\begin{tabular}{rlccccclrrrr}
\hline
\vspace{-0.3cm}
\\
\vspace{0.1cm}
(1) & \ \ \ \ (2) & \ \ \ \ (3) & \ (4) & \ (5) & (6) \ & \ \ \ \ (7) & \ (8) &
\ (9) & \ \ \ \ (10)& (11) & (12) \\ 
\# & star       &  J  & H  & K  & L & M   & references  & 12\,$\mu$m & 25\,$\mu$m  & 60\,$\mu$m & 100\,$\mu$m \\ 
\hline
\vspace{-0.2cm}
\\

  1& AB Aur        & 6.10  &  5.10 &  4.40 &  3.30 &  2.90 & HI                    &   27.16  &      48.10  &     105.60  &     114.10  \\ 
  2& UX Ori        & 8.03  &  7.43 &  6.71 &  5.61 &  5.32 & DW,26.2.1986          &    2.68  &       3.69  &       2.85  & $\sim$3.76  \\ 
  3& HD\,36112      & 7.44  &  6.70 &  5.90 &  4.75 &  4.47 & MA                    &    5.59  &      12.59  &      27.98  &      18.95  \\ 
  4& HK Ori        & 9.52  &  8.38 &  7.29 &  5.87 &  5.10 & HI                    &    3.80  &       4.08  &    $<$1.64  &   $<$70.37  \\ 
  5& HD\,245185     & 9.34  &  8.87 &  8.26 &  7.46 &  6.23 & HI                    &    4.00  &       5.96  &       4.97  &       3.61  \\ 
  6&  V380 Ori     & 8.25  &  7.10 &  6.03 &  4.44 &  3.61 & J-L ST;  M ME         &    8.61  &       8.85  &   $<$75.90  &   $<$38.59  \\ 
  7& HD\,37357      & 8.31  &  7.88 &  7.24 &  6.33 &       & MA                    &    2.01  &       2.79  & $\sim$2.90  &$\sim$20.29  \\ 
  8& HD\,37806      & 7.38  &  6.68 &  5.77 &  4.23 &  3.79 & MA                    &   11.02  &       9.40  &    $<$5.18  &   $<$33.98  \\ 
  9& HD\,95881      & 7.50  &  6.79 &  5.87 &  4.28 &  3.71 & MA                    &    9.14  &       6.87  &       1.45  &    $<$1.09  \\ 
 10& HD\,97048      & 7.30  &  6.75 &  6.04 &  4.61 &  4.56 & HI                    &   14.49  &      40.34  &      69.91  &  $<$250.10  \\ 
 11& HD\,100453     & 6.97  &  6.32 &  5.52 &  4.20 &  3.79 & MA                    &    7.23  &      33.59  &      39.36  &      23.86  \\ 
 12& HD\,100546     & 6.43  &  5.88 &  5.20 &  4.15 &  3.80 & MA                    &   65.78  &     242.60  &     165.20  &      98.56  \\ 
 13& HD\,101412     & 8.70  &  8.24 &  7.25 &  5.81 &  5.08 & DW                    &    3.22  &       3.09  & $\sim$1.69  &$\sim$10.52  \\ 
 14& HD\,104237     & 5.75  &  5.14 &  4.42 &  3.05 &  2.58 & MA                    &   23.65  &      23.05  &      14.72  &       9.58  \\ 
 15& HD\,135344     & 7.44  &  6.72 &  5.96 &  4.76 &  4.69 & MA                    &    1.59  &       6.71  &      25.61  &      25.69  \\ 
 16& HD\,139614     & 7.75  &  7.34 &  6.76 &  5.68 &  5.49 & MA                    &    4.11  &      18.14  &      19.30  &      13.94  \\ 
 17& HD\,142666     & 7.34  &  6.72 &  6.04 &  4.97 &  4.69 & MA                    &    8.57  &      11.21  &       7.23  &       5.46  \\ 
 18& HD\,142527     & 6.65  &  5.94 &  5.20 &  3.89 &  3.50 & MA                    &   10.38  &      21.23  &     105.10  &      84.70  \\ 
 19& HD\,144432     & 7.21  &  6.69 &  6.14 &  5.14 &  4.90 & MA                    &    7.53  &       9.36  &       5.76  &       3.29  \\ 
 20& HD\,144668     & 5.83  &  5.18 &  4.38 &  3.08 &  2.54 & HI                    &   18.05  &      14.51  &$\sim$14.36  &   $<$63.25  \\ 
 21& HD\,150193     & 7.05  &  6.37 &  5.64 &  4.37 &  3.93 & MA                    &   17.61  &      18.10  &       8.13  &   $<$16.25  \\ 
 22& HD\,163296     & 6.24  &  5.52 &  4.70 &  3.52 &  3.14 & MA                    &   18.20  &      20.99  &      28.24  &   $<$40.62  \\ 
 23& HD\,169142     & 7.43  &  7.01 &  6.53 &  5.64 &  5.57 & MA                    &    2.95  &      18.43  &      29.57  &      23.42  \\ 
 24& HD\,179218     & 6.99  &  6.64 &  5.91 &  4.68 &  4.18 & J,H EI; K-M LA        &   23.44  &      43.63  &      29.92  &      17.35  \\ 
\hline
\vspace{-0.3cm}
\end{tabular}
\end{center}
\end{table*}

\begin{figure}[t]
  \centerline{
  \hspace{-0.6cm}
  \includegraphics[width=9.4cm]{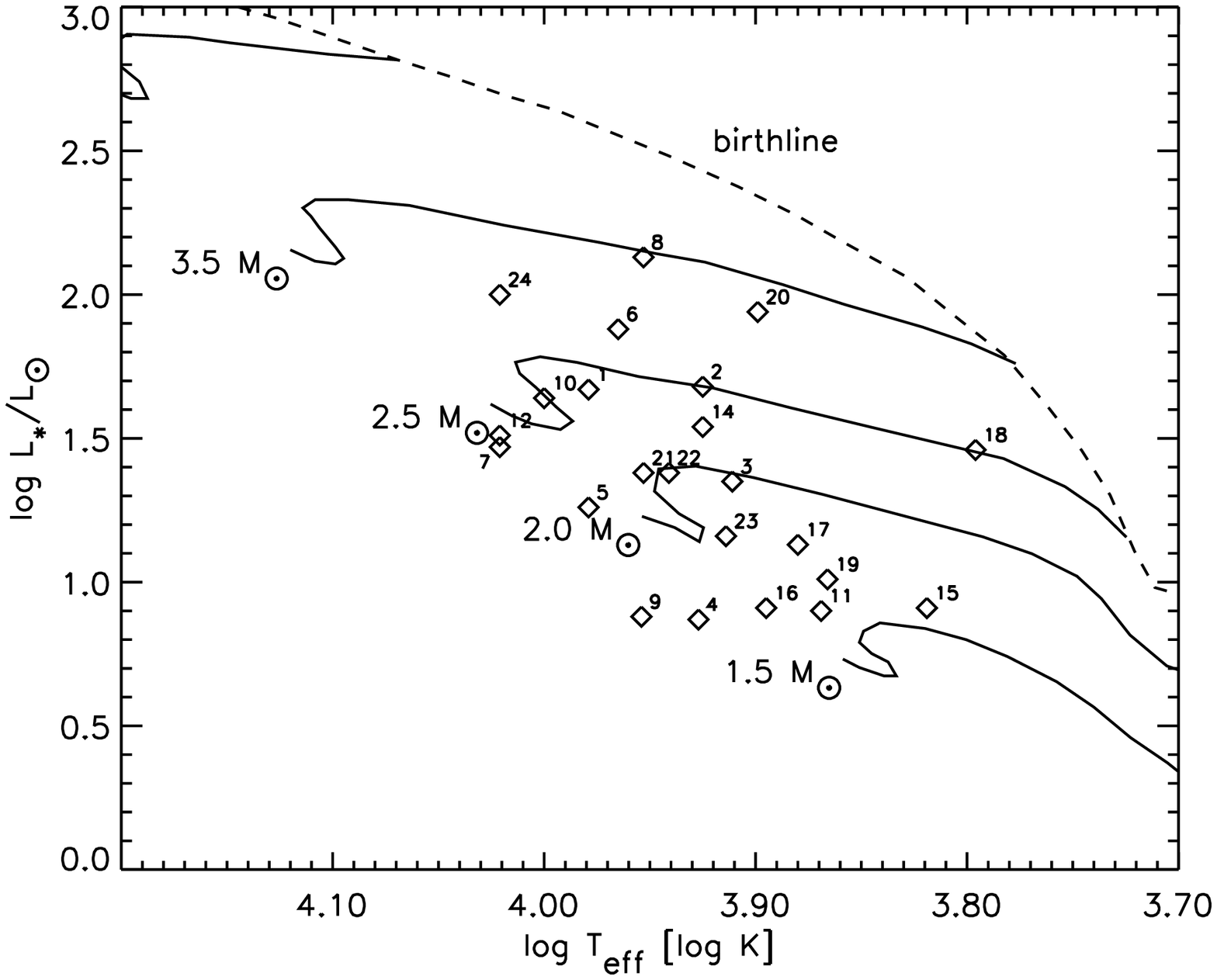}}

  \caption{ The positions of our stars in the HR diagram. The solid
curves indicate the pre-main sequence evolutionary tracks of stars of
different masses; the dashed curve represents the birthline for an
accretion rate of 10$^{-5}$~\msun yr$^{-1}$ \citep[both are taken
from][]{1993ApJ...418..414P}.  }
  \label{fig:t2_fs:HR_diagram}
\end{figure}

Our sample of stars was selected from a larger list of (candidate)
Herbig Ae/Be stars studied by \cite{1998A&A...331..211M}.  This list was
constructed by comparing the positions of stars in the Smithsonian
Astrophysical Observatory Star Catalog with positions in the Infrared
Astronomical Satellite (IRAS) point source catalogue, and subsequent
follow-up studies to find the Herbig star candidates. Clearly, such a way to
select stars may introduce biases. Our sample does not contain stars
that are heavily obscured due to e.g. on-going accretion or
an edge-on dusty disk. Instead, our sample is dominated by stars with
a relatively ``clean'' environment, low optical extinction, and disk
orientations that are not close to edge-on.
An overview of the stars in the sample and their basic parameters
is given in Table\,\ref{tab:t2_fs:source_list}.

In order to select ``genuine'' HAe stars, we applied the following
selection criteria:
\begin{enumerate}
\item
The position in the HR diagram should be in agreement with that of a
pre-main-sequence (PMS) star with a mass between 1.5 and 3.5\,\msun \
(see Fig.\,\ref{fig:t2_fs:HR_diagram}).

\item The color criterion J-H$>$0.25\,mag, assuring that the onset of the near-infrared
(NIR) excess is well defined, in agreement with the inner boundary of
the gas rich disk being set by the silicate evaporation
temperature of $\sim$1500\,K (see
Table\,\ref{tab:t2_fs:literature_photometry} for an overview of the
photometric data used).
\end{enumerate}

\subsection{Selection effects}

\begin{figure}[t]
  \centerline{
  \hspace{-0.6cm}
  \includegraphics[width=9.4cm]{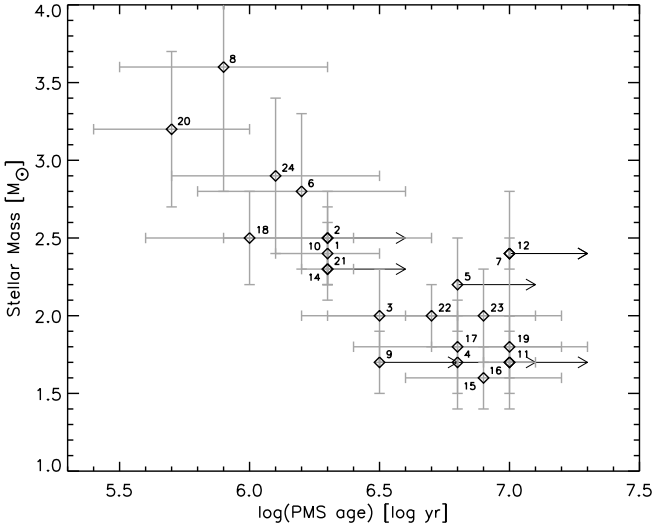}}
  \caption{The derived stellar mass estimates vs.
the pre-main-sequence age estimates derived from a comparison of their position
in the HR diagram to PMS evolutionary tracks of Palla \& Stahler (1993).
Our sample shows a clear lack of disks around relatively old
stars of 3-4\,\msun, and of disks around relatively young stars less massive than about 2.5\,\msun.}
  \label{fig:t2_fs:Mstar_vs_age}
\end{figure}

In order to identify if other selection effects may have entered our
sample, we determined the mass and age (calculated from the birthline,
for an accretion rate of 10$^{-5}$\,\msunyr)
of the stars by placing them in
the HR diagram (Fig.\,\ref{fig:t2_fs:HR_diagram})
and comparing their positions to
PMS evolutionary tracks published by \cite{1993ApJ...418..414P}.
Throughout this work ``(PMS) age'' $\tau$ is defined 
as the time that past since the star was on the 
birthline, which is when the star becomes optically
visible for the first time.
The luminosity is calculated from the observed photometry and the
measured distance to each star.
The uncertainties in the derived stellar masses and ages listed in
Table\,\ref{tab:t2_fs:source_list} and shown in Fig.~\ref{fig:t2_fs:Mstar_vs_age} mainly reflect the uncertainty in the distances \citep[see also][]{1998A&A...330..145V}. 
A comparison of the stellar parameters derived using the
\cite{1993ApJ...418..414P} evolutionary tracks to those obtained using more
recent calculations \citep{1999ApJ...525..772P,2000A&A...358..593S}, shows that
the particular choice of PMS tracks introduces an additional uncertainty in the
stellar ages that is of the same order as those implied by observational uncertainties in the position of the star in the HR diagram. This additional uncertainty is not taken into account.
In Fig.\,\ref{fig:t2_fs:Mstar_vs_age} we show the derived stellar
masses versus the PMS ages of the stars.
It is evident from this figure that our sample lacks "old" ($\tau \gtrsim 10^{6.3}$yr)
disks around 3-4\,M$_{\odot}$ stars, and  "young" ($\tau \lesssim 10^{6}$yr)
disks around stars less massive than about 2.5\,M$_{\odot}$. 
The lack of ``old'' disks around 3-4\,M$_{\odot}$ stars is likely
caused by the shorter timescale on which the disks around more
massive stars are dispersed. It is therefore not a selection effect.
The lack of ``young'' disks around the lower mass stars is
most likely
caused by the fact that these stars do not clear their environment
as rapidly as the more massive (more luminous) stars, and 
consequently become optically bright later in their evolution.
Also, at a given age, lower mass PMS stars are less
luminous than more massive ones, and so more easily escape optical
detection.
This selection effect must be taken into account in any discussion
about the evolution of the dust in proto-planetary disks based on
optically selected samples.

\subsection{Classification of the sources}
\label{sec:t2_fs:meeus_classification}

\begin{figure}[t]
  \centerline{
  \hspace{-0.6cm}
  \includegraphics[height=9.7cm,angle=90]{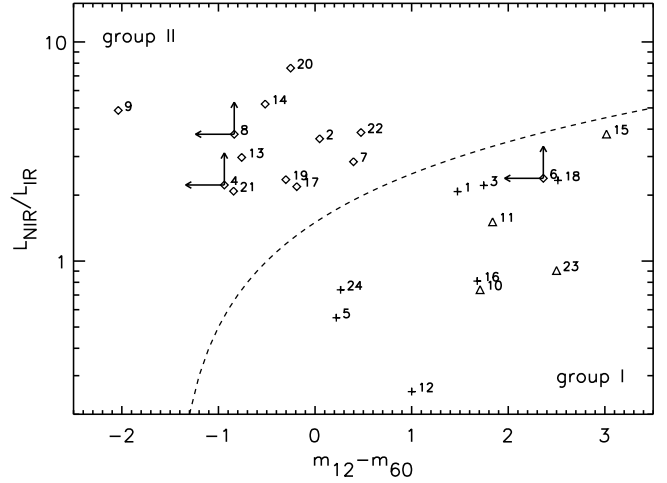}}
  \caption{Classification of the sources based on global SED properties.
We plot the ratio of the near-infrared and infrared luminosity
(see section~\ref{sec:t2_fs:meeus_classification}) vs. the
IRAS m$_{12}$-m$_{60}$ color (defined as
m$_{12}$-m$_{60}$=$-2.5 \log \mathcal{F}_{12}/\mathcal{F}_{60}$, where
$\mathcal{F}_{12}$ and $\mathcal{F}_{60}$ are the fluxes at 12 and
60\,$\mu$m as listed in the IRAS point source catalogue).
Sources classified as group\,I according to ME01 are in the
lower right part of the diagram, group\,II sources are in the upper
left part. Group\,Ia sources are indicated with crosses, group\,Ib
sources by triangles, and group\,II sources with diamonds.
The dashed curve indicates our division line between the two groups
(section~\ref{sec:t2_fs:meeus_classification}).}
  \label{fig:t2_fs:iras_color_NIRoverIR2}
\end{figure}

ME01 empirically decomposed the infrared spectra of Herbig Ae/Be stars into
three components: a power law component, a cold black-body component,
and solid state emission bands (mainly at 10 and 20\,$\mu$m).  They
found that some sources exhibit both the power law and cold black-body
component, and classified these sources as ``group\,I''.  Sources that
lack the cold black-body component were classified as ``group\,II''. A
further division into subgroups ``a'' and ``b'' serves to indicate the
presence or absence, respectively, of silicate emission bands at 10
and 20\,$\mu$m. It was proposed by ME01 that group\,I sources have a
large (several hundred AU) flared outer disk, whereas the group\,II
sources have a smaller, non-flaring outer disk.

We classify the sources for which we have newly measured N-band
(8-13.5$\mu$m)
spectra following ME01. Whereas ME01 had ISO spectra of their sources
at their disposal, our classification is based solely on broad-band
photometry. We find that the group\,I and group\,II sources are well
separated in an IRAS m$_{12}$-m$_{60}$ color versus $\mathrm
L_{\mathrm{NIR}}/\mathrm L_{\mathrm{IR}}$ diagram
(Fig.~\ref{fig:t2_fs:iras_color_NIRoverIR2}), where $\mathrm
L_{\mathrm{NIR}}$ is the integrated luminosity as derived from the
J,H,K,L and M band photometry, and $\mathrm L_{\mathrm{IR}}$ is the
corresponding quantity derived from the IRAS 12, 25 and 60\,$\mu$m
points.  For group\,I sources, $\mathrm L_{\mathrm{NIR}}/\mathrm
L_{\mathrm{IR}} \leq$ (m$_{12}$-m$_{60}$)+1.5, group\,II sources have
$\mathrm L_{\mathrm{NIR}}/\mathrm L_{\mathrm{IR}} >$
(m$_{12}$-m$_{60}$)+1.5. The dashed line in
Fig.\,\ref{fig:t2_fs:iras_color_NIRoverIR2} indicates the boundary
between the two groups.
To avoid any pre-assumptions concerning the spectral shapes of our targets, we
did not apply color correction to the IRAS data for the classification of the
sources.

By use of the
$L_{\mathrm{NIR}}$/$L_{\mathrm{IR}}$ ratio
\citep[][]{2003A&A...400L..21V}
we compare the near-IR SED to the mid-IR SED.
The near-IR SED is found to be similar for all HAe stars
\citep[][]{2001A&A...371..186N}, while the major differences occur in
the mid-IR SED. The $L_{\mathrm{NIR}}/L_{\mathrm{IR}}$ 
ratio is smaller for group\,I than
for group\,II sources. The mid-IR SED of group\,I sources
is ``double-peaked'' compared to the SED of a group\,II source.  
Group\,I sources are redder than their group\,II counterparts. The IRAS
m$_{12}$-m$_{60}$ color index quantifies this difference in SED shape.

\section{Observations and data reduction}
\label{sec:t2_fs:observations_data_reduction}

\begin{table*}

\begin{center}

\label{tab:t2_fs:observing_log}
\caption{Log of the TIMMI\,2 observations. We list the observing date
(defined as the day on which each observing night began), time (U.T.),
airmass of the observation, and integration time in seconds (columns\,3-6). The
calibrators used for the atmospheric correction are also given, with
the time and airmass of the measurements (columns\,7-12).}

\begin{tabular}{rllllrllllll}
\hline
\vspace{-0.3cm}
\\
\vspace{0.1cm}
(1) &\ \ \ \ (2) & \ \ \ \ (3) & \ (4) & \ (5) & (6) \ & \ \ \ \ (7) & \ (8) &
\ (9) & \ \ \ \ (10) & (11) & (12) \\
\# & star       &  date       & time  &$m_{\rm A}$& T$_\mathrm{int}$ & calibrator 1 & time &$m_{\rm A,1}$  & calibrator 2 & time  & $m_{A,2}$ \\
\hline
\vspace{-0.2cm}
\\
 2&      UX Ori &  27-12-2001 & 01:09 & 1.31 & 1380  &  HD\,32887 & 01:59 & 1.06  &  HD\,32887     & 00:39 & 1.25 \\
 3&    HD\,36112 &  18-03-2003 & 23:19 & 1.83 &  960  &  HD\,48915 & 00:27 & 1.04  &  HD\,29139     & 23:58 & 1.88 \\
 4&      HK Ori &  19-03-2003 & 23:21 & 1.39 &  960  &  HD\,48915 & 00:56 & 1.08  &  HD\,23249     & 00:04 & 1.68 \\
 5&   HD\,245185 &  19-03-2003 & 00:27 & 1.56 &  720  &  HD\,48915 & 00:56 & 1.08  &  HD\,23249     & 00:04 & 1.68 \\
 6&    V380 Ori &  17-03-2003 & 00:59 & 1.37 &  960  &  HD\,48915 & 00:21 & 1.04  &  HD\,58972     & 01:36 & 1.34 \\
 7&    HD\,37357 &  19-03-2003 & 01:08 & 1.43 &  720  &  HD\,48915 & 00:56 & 1.08  &  HD\,55865     & 02:33 & 1.46 \\
 8&    HD\,37806 &  27-12-2001 & 04:04 & 1.12 &  690  &  HD\,32887 & 01:59 & 1.06  &  HD\,32887     & 00:39 & 1.25 \\
 9&    HD\,95881 &  17-03-2003 & 02:55 & 1.37 &  720  &  HD\,48915 & 00:21 & 1.04  & HD\,107446     & 02:41 & 1.32 \\
10&    HD\,97048 &  19-03-2003 & 03:02 & 1.52 &  720  &  HD\,98292 & 04:02 & 1.28  &  HD\,55865     & 02:33 & 1.46 \\
11&   HD\,100453 &  18-03-2003 & 05:24 & 1.13 &  960  &  HD\,93813 & 05:00 & 1.08  & HD\,146003     & 06:38 & 1.25 \\
12&   HD\,100546 &  19-03-2003 & 05:09 & 1.34 &  960  & HD\,109379 & 05:36 & 1.01  &  HD\,98292     & 06:18 & 1.37 \\
13&   HD\,101412 &  17-03-2003 & 03:25 & 1.19 &  960  &  HD\,48915 & 00:21 & 1.04  &  HD\,89388     & 04:04 & 1.19 \\
14&   HD\,104237 &  17-03-2003 & 04:57 & 1.52 &  480  & HD\,123139 & 08:04 & 1.03  &  HD\,89484     & 04:37 & 1.65 \\
15&   HD\,135344 &  17-03-2003 & 05:56 & 1.13 &  960  & HD\,123139 & 08:04 & 1.03  & HD\,107446     & 06:35 & 1.20 \\
16&   HD\,139614 &  17-03-2003 & 07:25 & 1.05 &  960  & HD\,123139 & 08:04 & 1.03  & HD\,107446     & 06:35 & 1.20 \\
17&   HD\,142666 &  18-03-2003 & 06:07 & 1.23 &  720  &  HD\,93813 & 05:00 & 1.08  & HD\,146003     & 06:38 & 1.25 \\
18&   HD\,142527 &  18-03-2003 & 06:56 & 1.11 &  720  & HD\,123139 & 08:07 & 1.04  & HD\,146003     & 06:38 & 1.25 \\
19&   HD\,144432 &  18-03-2003 & 07:25 & 1.06 &  960  & HD\,123139 & 08:07 & 1.04  & HD\,146003     & 06:38 & 1.25 \\
20&   HD\,144668 &  17-03-2003 & 09:15 & 1.02 &  720  & HD\,152334 & 09:45 & 1.03  & HD\,152786     & 09:01 & 1.14 \\
21&   HD\,150193 &  18-03-2003 & 08:23 & 1.04 &  720  & HD\,123139 & 08:07 & 1.04  & HD\,146003     & 06:38 & 1.25 \\
22&   HD\,163296 &  18-03-2003 & 09:40 & 1.04 &  480  & HD\,123139 & 08:07 & 1.04  & HD\,152786     & 09:22 & 1.12 \\
23&   HD\,169142 &  17-03-2003 & 10:07 & 1.03 &  960  & HD\,152334 & 09:45 & 1.03  & HD\,152786     & 09:01 & 1.14 \\
24&   HD\,179218 &  19-03-2003 & 09:39 & 1.78 &  480  & HD\,152334 & 07:59 & 1.10  & HD\,187642     & 10:00 & 1.68 \\

\hline
\vspace{-0.3cm}
\end{tabular} \\
\end{center}
\end{table*}

Infrared spectra in the 10\,$\mu$m atmospheric window were taken in
December 2001 and March 2003 with the TIMMI2 instrument
mounted at the 3.60\,m telescope at ESO La Silla observatory.
Conditions were clear during all nights.  The low resolution
(R$\approx$160) N~band grism was used in combination with a 1.2
arcsecond slit. The pixel scale in the spectroscopic mode of TIMMI2 is
0.45 arcseconds. A log of our observations is given in 
Table\,\ref{tab:t2_fs:observing_log}.

\subsection{Atmospheric correction}
\label{sec:t2_fs:atmospheric_correction}
Ground-based observations at 10\,$\mu$m are particularly challenging
because of the high atmospheric and instrumental background, and the
varying transmission of the earth atmosphere. As proper spectral
calibration is essential for the compositional analysis presented in
section~\ref{sec:t2_fs:compositional_fits}, we report here in some detail on
our data reduction method.

To deal with the high background signal, we employed standard chopping and
nodding, using a $+10$ arcsecond chop throw North-South, and a $-10$
arcsecond nod throw North-South.
For the spectral calibration of our measurements we regularly observed standard
stars. These observations are used to determine both the
atmospheric extinction per unit airmass ($A_{\nu}$) and the instrumental 
response ($R_{\nu}$) at all wavelengths. For each science observation, we
determine $A_{\nu}$ and $R_{\nu}$ from two calibration measurements:

\begin{align}
s_{\nu,1} & =I_{\nu,1}\,e^{-\tau_{\nu,1}}R_{\nu}  ; &
\tau_{\nu,1} &=A_{\nu} m_{\rm A,1} \\
\label{eq:t2_fs:tau}
\nonumber
s_{\nu,2} & =I_{\nu,2}\,e^{-\tau_{\nu,2}}R_{\nu}  ; &
\tau_{\nu,2} & =A_{\nu} m_{\rm A,2},
\end{align}

\noindent
where $\mathrm{s_{\nu,i}}$ are the measured calibrator spectra,
$\mathrm{I_{\nu,i}}$ are the intrinsic (model) calibrator spectra,
$\tau_{\nu,i}$ are the optical depths of the earth atmosphere 
during the calibration measurements, and $m_{\rm{A},i}$
denote the airmass at which the calibrators were observed.  Whenever
possible, we chose $m_{\rm{A,1}}$ or $m_{\rm{A,2}}$
to be very close to the airmass
of the science observation. Both calibrator measurements are chosen as
close in time as possible to the science observation.  For the
intrinsic calibrator spectra we use ``spectral templates'' by
\cite{1999AJ....117.1864C}\footnote{available for download at the
TIMMI2 website
(http://www.ls.eso.org/lasilla/sciops/timmi/docs/tables/)}.
Table\,\ref{tab:t2_fs:calibrator_templates} lists the observed
calibrators and the applied templates, which are chosen to match the
calibrator spectral type as closely as possible. Solving 
Eq.\,1
for $A_{\nu}$ and $R_{\nu}$, we find:

\begin{equation}
A_{\nu}=\frac{\ln(I_{\nu,2}/I_{\nu,1})+\ln(s_{\nu,1}/s_{\nu,2})}
{m_{\rm A,2}-m_{\rm A,1}}
\end{equation}

\begin{equation}
R_{\nu}=\frac{s_{\nu,1}}{I_{\nu,1}e^{-A_{\nu} m_{\rm A,1}}}
\end{equation}
The intrinsic spectrum of a science target observed at airmass $m_{\rm A}$
is then calculated from its measured spectrum $s_{\nu}$ as:
\begin{equation}
I_{\nu}=\frac{s_{\nu}}{R_{\nu}}e^{A_{\nu} m_{rm A}}
\end{equation}
For a number of sources we have ISO spectra available. We generally
find very good agreement in spectral shape between the ISO data and
our new ground based spectra. A comparison with the ISO spectra of the
brighter stars (HD\,100546, HD\,163296, HD\,150193) shows differences
at the level of at most a few percent in the shape of the spectra, when
the TIMMI2 spectra are scaled such as to most closely match the flux 
levels in the ISO spectra. For each HAe star spectrum, the time and airmass 
of observation, as well as the time and airmass of the two calibration
observations are given in Table\,\ref{tab:t2_fs:observing_log}.

\subsection{Flux calibration}
\label{sec:t2_fs:flux_calibration}
Our stars have been selected to be \emph{isolated} Herbig stars. Therefore,
the IRAS photometry will in most cases not be contaminated by emission from 
nearby sources or a surrounding remnant cloud. All emission seen
in the IRAS data is expected to originate in the disk.
We flux-calibrated the spectra using the IRAS 12\,$\mu$m data, by
applying a scaling factor $q$ such that
\begin{equation}
q\int_{\nu=0}^{\infty} I_{\nu}T_{\nu}d\nu =  \mathcal{F}_{12} \times 1.35 \times 10^{-10}
\ \ \ \mathrm{[erg \ s^{-1} cm^{-2}]},
\end{equation}
where $T_{\nu}$ is the normalized instrumental response function of
the IRAS 12 micron band, and $\mathcal{F}_{12}$ is the 12 micron flux
as listed in the IRAS point source catalogue
(a source that has a 12\,$\mu$m flux of $\mathcal{F}_{12}$=1\,Jy in 
the IRAS point source
catalogue, yields an inband flux of 
$1.35\times 10^{-10}$ ${\rm erg \ s^{-1} cm^{-2}}$). The IRAS 12 micron band
runs from 8 to 15\,$\mu$m and is therefore somewhat broader than our
spectral coverage. To allow for the calibration we estimate the
spectrum between 13.5 and 15\,$\mu$m, by linearly extrapolating $I_{\nu}$
between the
continuum points measured at 8 and 13\,$\mu$m. Our final calibrated
spectrum is then $F_{\nu}=qI_{\nu}$. We estimate the absolute
photometric accuracy to be 15\%.

\begin{table}[!t]
\begin{center}
\caption{Spectral templates used for the calibrators
(see section~\ref{sec:t2_fs:atmospheric_correction}).}
\begin{tabular}{llll}
\hline
calibrator & spectral   & template file  & template \\
           & type       &                & sp. type \\
\hline
HD\,23249   & K0IV       & hd123139.tem       & K0IIIb      \\
HD\,29139   & K5III      & alp\_Tau.dat       & K5III       \\
HD\,48915   & A1V        & $\nu^2$ law        & $\nu^2$ law \\
HD\,55865   & K0III      & hd123139.tem       & K0IIIb      \\
HD\,58972   & K3III      & hd6805.tem         & K2III       \\
HD\,89388   & K3IIa      & hd6805.tem         & K2III       \\
HD\,89484   & K1IIIb     & hd169916.tem       & K1IIIb      \\
HD\,92397   & K4.5III    & hd32887.tem        & K4III       \\
HD\,93813   & K0/K1III   & hd123139.tem       & K0IIIb      \\
HD\,98292   & M2III      & alp\_Tau.dat       & K5III       \\
HD\,107446  & K3.5III    & hd32887.tem        & K4III       \\
HD\,109379  & G5II       & hd37160.tem        & G8III-IV    \\
HD\,123139  & K0IIIb     & hd123139.tem       & K0IIIb      \\
HD\,146003  & M2III      & alp\_Tau.dat       & K5III       \\
HD\,152334  & K4III      & hd32887.tem        & K4III       \\
HD\,152786  & K3III2     & hd6805.tem         & K2III       \\
HD\,177716  & K1IIIb     & hd169916.tem       & K1IIIb      \\
HD\,187642  & A7V        & HD187642\_spec.dat & A7IV-V      \\
\hline
\end{tabular}
\label{tab:t2_fs:calibrator_templates}
\end{center}
\end{table}

\section{Description of the observations}
\label{sec:t2_fs:description}

\subsection{Description of the spectra}
The spectra of our sample of HAe stars are shown in
Fig.~\ref{fig:t2_fs:all_spectra}. All stars show spectral structure on
top of a continuum whose slope varies strongly from star to star. The
spectral features are due to various kinds of silicates (see
section~\ref{sec:t2_fs:compositional_fits} below). Also, emission from
Polycyclic Aromatic Hydrocarbons (PAHs) can be seen, at 7.9, 8.6, 11.3
and 12.7\,$\mu$m.  The relative importance of the silicate and PAH
contributions varies strongly.  There are sources that show both
silicate and PAH emission (e.g. HD\,100546, HD\,179218), sources that show only
silicate emission (e.g. HD\,144432), and sources that display only
PAH emission (these are the group\,Ib sources HD\,97048, HD\,100453,
HD\,135344, and HD\,169142).
There are no sources in our sample
that have a completely featureless 10~micron spectrum.

ME01 and \cite{2004A&A...426..151A} found that the 
SED correlates with the
presence and/or strength of the PAH bands: group\,I sources tend to show
(prominent) PAH emission, while group\,II sources do not. This trend is
confirmed in our sample, but we note that there is considerable
scatter. For instance, HD\,95881 has little far-IR excess and is
thus classified as group\,II,  but, nonetheless, shows clear PAH emission bands.

The silicate band shows very large variations in shape and strength.
The bulk of the emission is in most cases due to amorphous silicates,
but almost all stars show some spectral structure near 11.3\,$\mu$m,
which can be attributed to forsterite. Note, however, that this
feature blends with the 11.3\,$\mu$m PAH band. There are also prominent narrow
emission bands near 9.2 and 10.6\,$\mu$m. These are due to crystalline 
enstatite. The spectrum of HD\,100546 is
dominated by crystalline forsterite, while that of HD\,179218 is dominated
by crystalline enstatite. This latter star shows one of the richest
10\,$\mu$m spectra observed to date (see
Fig.\ref{fig:t2_fs:HD179218_enstatite}). The resonances of crystalline
enstatite are clearly visible in the spectrum of this source.
The ISO spectrum of
HD\,179218 at longer wavelengths
also points to a relatively high abundance of crystalline
enstatite \citep{2001A&A...375..950B}.  It is obvious that the nature
of the crystalline dust in our sample shows very large variations,
both in terms of the fraction of the dust that is crystalline, and in
composition.

There are four stars (HD\,97048, HD\,100453, HD\,135344 and HD\,169142),
all classified as group\,I, that show no detectable silicate emission.
Instead, their 10\,$\mu$m spectra are dominated by PAH emission. The
lack of silicate emission is most simply explained by assuming that
there are no small ($<$~3-5\,$\mu$m) silicate grains in the inner
10-20\,AU of the disk. \cite{2002A&A...392.1039M} derive limits on the
presence of small silicate grains in HD\,100453, and argue that all
grains smaller than 4\,$\mu$m must have been removed. The most likely
cause for the removal of small silicate grains is grain growth, but
apparently this has not affected the population of small carbonaceous
grains to the same extent \citep{2004A&A...418..177V}.
Possibly, small grains
survive in the outer disk regions. At large distance from the star,
the silicate grains may be too cold to contribute to the 10\,$\mu$m
spectrum, while the PAHs can still produce significant emission. 
This can only occur if the PAHs have not been incorporated into
larger grains. Indeed, \cite{2004Natur.432..479V}
find evidence for a distance dependence of the typical silicate grain
size in the surface layers of HAe star disks: in the innermost regions
growth has proceeded further than in the outer disk regions.

\begin{figure*}[!t]
  \centerline{ 
 \resizebox{\hsize}{!}{\rotatebox{90}{\includegraphics{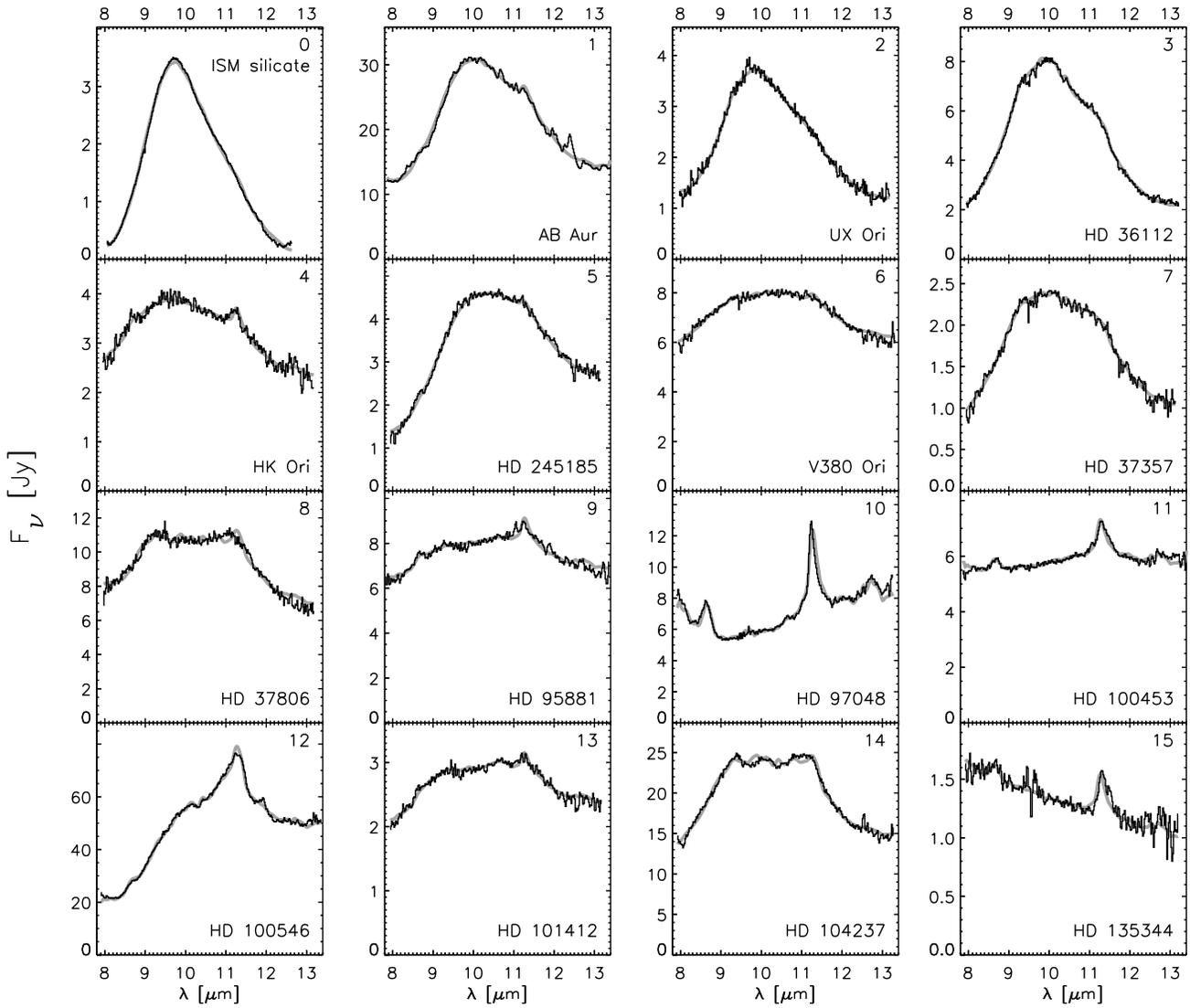}}}}
  \caption{N-band spectra of the sources in our sample. 
The ISM silicate extinction efficiency, plotted in the upper left panel,
was taken from \cite{2004ApJ...609..826K}. The
AB~Aur spectrum was taken by ISO \citep{2000A&A...357..325V}.
Also plotted are the best fits to the spectra (grey curves,
see section\,\ref{sec:t2_fs:results}).}
  \label{fig:t2_fs:all_spectra}
\end{figure*}

\setcounter{figure}{3}
\begin{figure*}[!t]
  \centerline{
 \resizebox{\hsize}{!}{\rotatebox{90}{\includegraphics{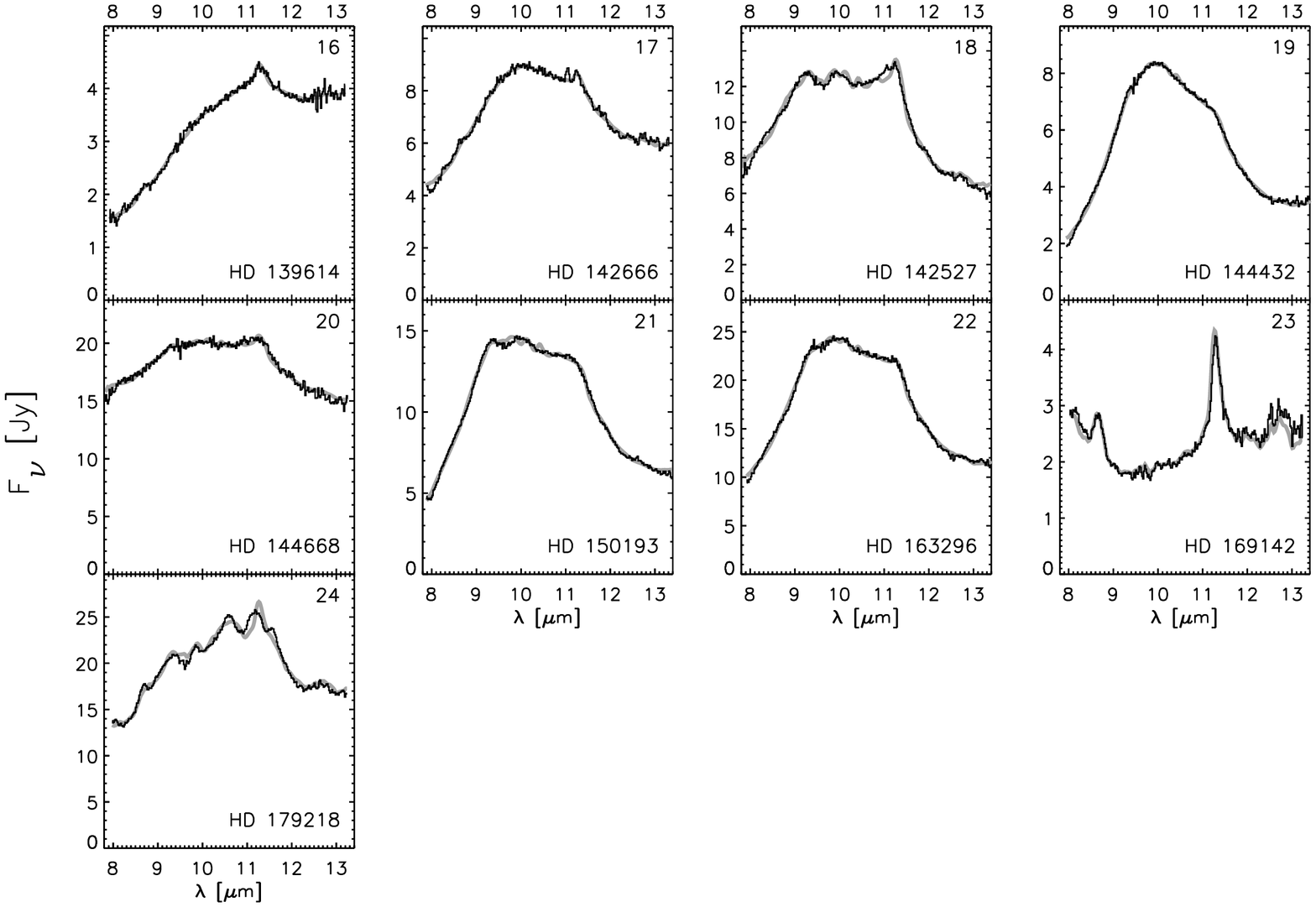}}}}
  \caption[]{(Continued)}
\end{figure*}

\begin{figure}[t]
  \centerline{  
  \hspace{-0.22cm}
  \includegraphics[width=9.1cm,angle=0]{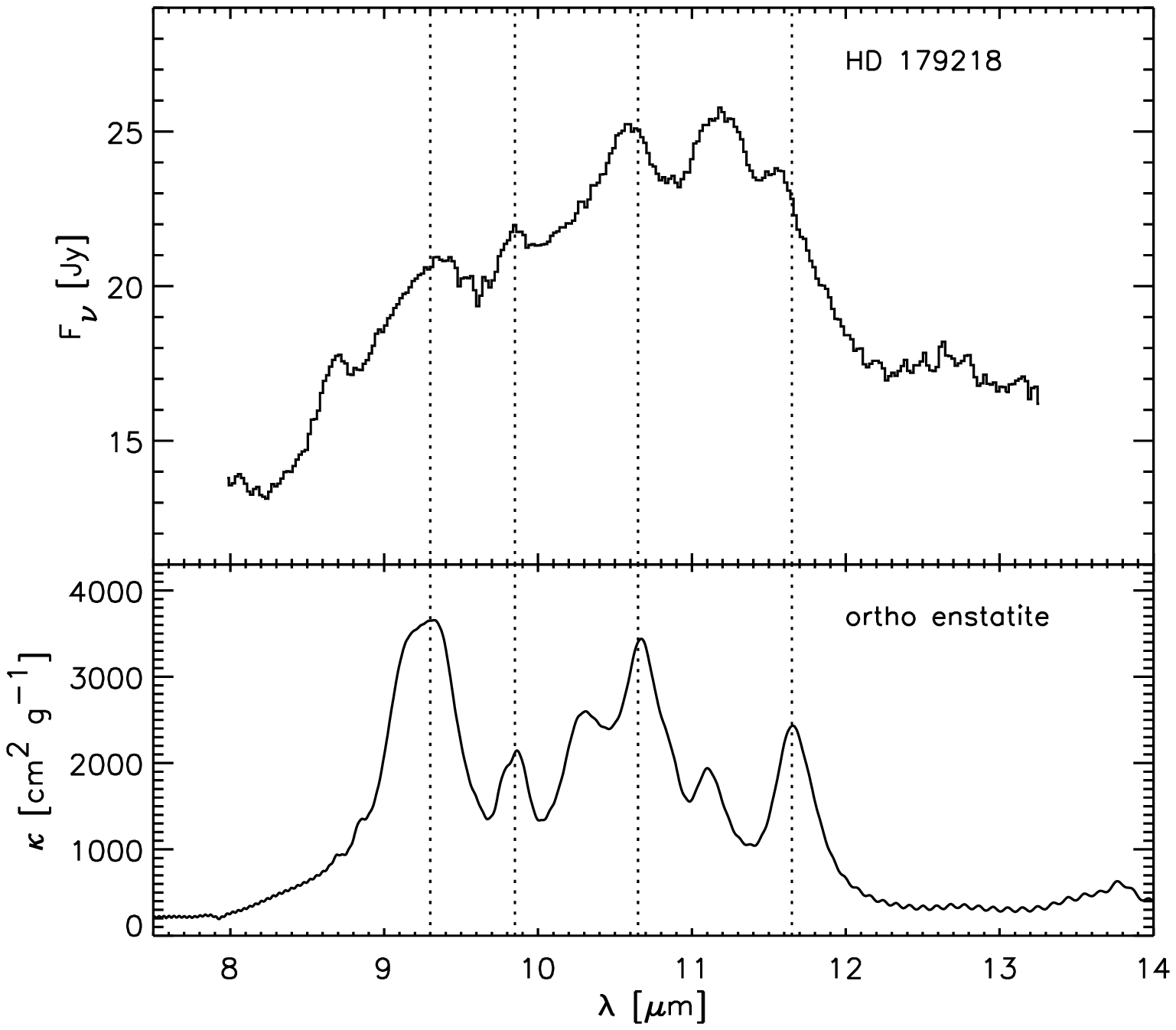}}
  \caption{
The N-band spectrum of HD\,179218 (upper panel), and the measured mass
absorption coefficients of ortho enstatite taken from
\cite{2002A&A...391..267C} (lower panel).  The wavelengths of the most
prominent emission bands are indicated by the dotted lines.
In this object, enstatite grains are an important constituent
of the grain population that causes the 10\,$\mu$m feature.}
  \label{fig:t2_fs:HD179218_enstatite}
\end{figure}

\subsection{The shape and strength of the silicate feature}

\begin{figure}[t]
  \hspace{-0.2cm}
  \centerline{
  \includegraphics[width=9.6cm]{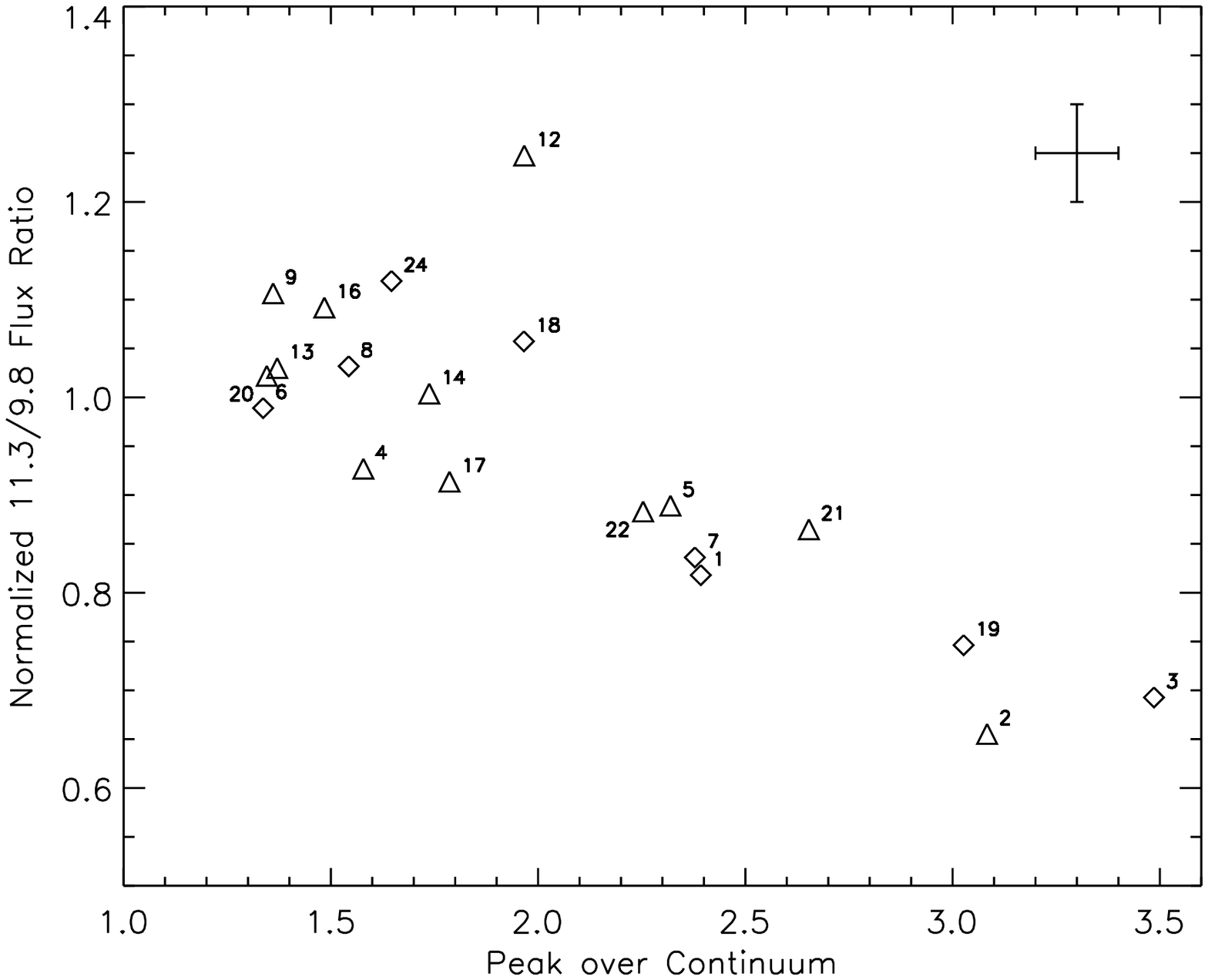}}
  \caption{The flux ratio of the normalized spectra
at 11.3 and 9.8 micron 
(a measure for the amount of processing
that the material has undergone)
versus the peak/continuum ratio of the silicate
feature (a measure for the typical grain size).
Group\,I sources are represented by triangles, group\,II sources
by diamonds.
In the upper right corner of the figure we have indicated
the typical uncertainties in the displayed quantities.}
  \label{fig:t2_fs:feature_cc_all_LinearContinuum}
\end{figure}

In Fig.\,\ref{fig:t2_fs:feature_cc_all_LinearContinuum} we show the
ratio of the normalized spectrum at 11.3 and 9.8\,$\mu$m, 
against the silicate peak to continuum ratio 
(``feature strength'')\footnote{
To estimate the continuum we simply interpolate linearly 
between 8 and 13 micron. The peak/continuum ratio is the maximum
value of the normalized spectrum 
$F_{norm}=1+F_{\nu,cs}/<F_{\nu,c}>$,
where $F_{\nu,cs}$ is the continuum subtracted spectrum 
($F_{\nu}-F_{\nu,c}$) and
$<F_{\nu,c}>$ is the mean of the continuum.
This definition of $F_{norm}$ 
preserves the shape of the emission band even if the continuum is not 
constant. For a constant continuum level it is identical to 
$F_{\nu}/F_{\nu,c}$.}.
Group\,I sources are indicated with triangles,
group\,II sources with diamonds (the ``outlier'' \#\,12 is
HD\,100546).  Sources with a low 11.3/9.8 ratio have a triangular
shaped emission feature, clearly peaked just shortward of 10\,$\mu$m
(e.g. UX~Ori, in Fig.\,\ref{fig:t2_fs:all_spectra}).  Sources
with a high 11.3/9.8 generally have a broad, flat-topped emission
band, often showing substructure (e.g. HD\,37806, HD\,142527).  There is
a clear correlation between the shape and strength of the silicate
feature; stars with a strong feature (i.e. a high peak/continuum
ratio) have a low 11.3/9.8 ratio, whereas stars with weaker silicate
features have higher 11.3/9.8 ratios. This correlation was first
demonstrated in Herbig Ae stars by \cite{2003A&A...400L..21V}.
\cite{2003A&A...409L..25M} and \cite{2003A&A...412L..43P} have
subsequently shown that the same trend is observed in the silicate
feature of T-Tauri stars.

The shape of the emission bands with a low 11.3/9.8 ratio is similar
to that of the ISM silicate absorption feature. Such 10 micron features are
indicative of small, amorphous silicate grains, i.e. relatively ``primitive''
dust.  The emission bands with high 11.3/9.8 ratios can be explained
with on average larger grains, and a higher degree of crystallinity,
i.e. relatively ``processed'' dust.  Thus, the silicate feature
11.3/9.8 ratio is a measure of the amount of processing that the
material has undergone
\citep{2001A&A...375..950B}.

\section{Analysis}
\label{sec:t2_fs:analysis}

\subsection{Compositional fits}
\label{sec:t2_fs:compositional_fits}

To derive the composition of the silicate dust causing the 10~micron
feature, the observed spectra have been fitted using the most commonly found
dust species in circumstellar material that show spectral structure in 
the 10~micron region (see Fig.\,~\ref{fig:t2_fs:templates}).
These are amorphous and
crystalline olivine and pyroxene, and amorphous silica
\citep[e.g.][]{2001A&A...375..950B}. Amorphous
olivine (Mg$_{2x}$Fe$_{2(1-x)}$SiO$_4$, where $0 \lesssim x \lesssim 1$ 
denotes the magnesium
content of the material) 
is the most commonly found
silicate in astrophysical environments. It dominates the 10~micron
extinction caused by dust grains in the ISM
\citep{2004ApJ...609..826K}.  The 10~micron emission spectrum of small
amorphous olivine grains is characterized by a rather broad feature
which peaks at 9.8\,$\mu$m. Small amorphous pyroxene grains
(Mg$_x$Fe$_{1-x}$SiO$_3$) show an emission feature very similar to
that of amorphous olivine grains, though shifted toward shorter wavelengths.
The emission spectra from small crystalline olivine and pyroxene grains show
strong, narrow resonances that are observed in
for example circumstellar disks \citep{1998ARA&A..36..233W,
2001A&A...375..950B} and comets \citep{1997Sci...275.1904C,
2003A&A...401..577B}. From the positions of the resonances in the emission
spectra of these objects it is clear that the magnesium rich
components dominate the emission \citep[see
e.g.][]{Jaeger1998}. In our fitting procedure we therefore
use crystalline olivine and pyroxene with $x=1$, i.e. forsterite and
enstatite respectively. For the amorphous olivine and pyroxene we use
$x=0.5$.

Studies of interplanetary dust particles \citep{1989LPSC...19..513R}
have shown that some of these particles contain large inclusions of
silica (SiO$_2$). Also, from laboratory experiments it is suggested
that when amorphous silicates are annealed to form forsterite, silica
will be formed \citep[see e.g.][]{2000A&A...364..282F}.  The emission
spectrum of silica in the 10 micron region has a distinct spectral
signature, with a strong feature peaking at 8.9\,$\mu$m.  Therefore,
silica is included as one of the possible dust components.

Many of our sources show emission bands at 7.9, 8.6, 11.3 and
12.7\,$\mu$m that are attributed to PAHs.
In order to include the PAH emission in our compositional fits, we
constructed a simple PAH template. This was done by taking our two highest
quality spectra of sources without a silicate feature 
(HD\,97048 and HD\,169142),
subtracting the continuum emission, and averaging over the two spectra.
The resulting PAH emission spectrum was added as a fit component, and
is shown in the bottom panel of Fig.~\ref{fig:t2_fs:templates}.

Dust grains in circumstellar environments are most likely aggregates
of many different dust species. However, in order to perform an analysis of the
abundances of the various components in a large sample of sources, it is not
(yet) feasible to do computations for such complex aggregates. If the
aggregates are very fluffy, the constituents of which they are composed will
interact with the incident radiation as separate entities.
\citet{2003LPI....34.1148M} show that indeed a measurement of the infrared
spectrum of a relatively large IDP still displays the spectral structure one
would expect from much smaller particles. Therefore, we assume that the
emission properties of an aggregated structure can be represented by the sum of
the emission properties of the constituents \citep[for a similar approach see
e.g.][]{1999P&SS...47..773B, 2001A&A...375..950B, 2004ApJ...610L..49H}.

All dust grains contributing to the 10$\,\mu$m region will in principle have
their own temperature depending on the grain size and composition as well as on
position in the disk. Consequently, the emission function of the dust will be a
(weighted) sum over blackbody curves of different temperatures. In order to
account for these effects one should use a disk model that self-consistently
describes the disk geometry and all relevant radiation processes. This is
beyond the scope of the analysis presented here. For state of the art disk
models we refer to \citet{2004A&A...417..159D}. We add that even these
sophisticated models still do not account for many relevant processes,
including the settling and (turbulent) mixing of grains leading to spatial
gradients in the dust properties. Such spatial gradients have been reported by
\citet{2004Natur.432..479V} to exist in Herbig Ae disks. One should also realize
that even if such a detailed modeling approach was attempted, the modest
wavelength interval provided by the TIMMI2 data would reveal only very limited
information on the temperature distribution. The continuum in the 10 micron
region is a very smooth, almost linear function of wavelength. We therefore
opted to represent the continuum by a single blackbody curve with a
temperature, $T_c$ that is characteristic for the dust emission as a
whole. Inherent in this approach is that we essentially assume that the
temperature of the individual grains is independent of composition and size,
and that there are no gradients in grain properties (per unit mass) throughout
the disk.

The disk regions that we study are partially optically thick at
10$\,\mu$m. However, we assume that the disk surface layer from which the
observed flux originates is optically thin at this wavelength. This must be the
case since we see emission features.
The flux emitted by a distribution of dust grains is then given by
\begin{equation}
\mathcal{F}_\nu^\mathrm{\,silicate}\propto B_\nu(T_c)~\sum_i w_i \kappa_i,
\end{equation}
where $B_\nu(T_c)$ denotes the Planck function at the characteristic temperature $T_c$,
$\kappa_i$ is the mass absorption
coefficient of dust component $i$ (see also Fig.\,\ref{fig:t2_fs:templates}),
and $w_i$ is a
weighting factor which is proportional to the total dust mass in
component $i$. The summation is over all dust components. The mass
absorption coefficient of each dust species is determined by the size,
shape, structure and chemical composition of the dust grains. The total
model spectrum $\mathcal{F}_\nu^\mathrm{model}$ is then calculated by
adding a continuum and PAH contribution to the silicate emission.

\subsubsection{Shape of the dust grains}

The shape and structure of the dust grains are very important
parameters determining the feature shape of the emission
spectrum. Usually it is assumed that the grains are homogeneous and
spherical so that Mie theory can be applied to calculate the
$\kappa_i$. Another widely used assumption is that the grains are much
smaller than the wavelength of radiation (``Rayleigh limit'') in which
case it is mathematically straightforward to adopt a continuous
distribution of ellipsoids
\citep[CDE][]{BohrenHuffman,2003A&A...401..577B}.  
Since micron sized silicate grains are not in the Rayleigh limit at a 
wavelength of 10\,$\mu$m,  we cannot use CDE calculations
to study grain growth.
Furthermore, a comparison between calculations of the mass absorption
coefficients of small crystalline silicates with measurements shows
that we cannot get good agreement using homogeneous spherical
particles \citep[see for example][]{2001A&A...378..228F}.

Adopting different grain shapes, \cite{2003A&A...404...35M} showed that the
absorption properties can be divided in essentially two categories. One
category contains the perfect homogeneous spheres; the other all other
investigated shapes, including hollow spheres. Shape effects within the second
category do exist, but they are small compared to the differences with
homogeneous spheres. One could say that the difference between perfect
homogeneous particles and those having other shapes is essentially a result of
a breaking of perfect symmetry \citep[see also][]{2003A&A...404...35M}.  This
implies that we have only very limited information on the true (likely
irregular) shape of astrophysical dust grains from spectroscopic analysis.
However, a practical implication of this result is that one may represent the
absorption properties of irregular grains with sufficient accuracy by adopting
the average properties of a distribution of shapes other than that of
homogeneous spheres. For this purpose, a practical choice is a distribution of
hollow spheres (DHS),  simply averaging over the volume fraction occupied by
the central inclusion, which ranges from 0 to 1. In this shape distribution,
the material volume of the particle is kept constant, thus particles
with a high value of $f$ will have a large outer radius. This shape
distribution has the advantage that it can be applied for all grain sizes using
a simple extension of Mie theory. \cite{2003A&A...404...35M} showed that this
indeed gives excellent results for small forsterite grains. Therefore, in this
work, the mass absorption coefficients of all crystalline grains (forsterite,
enstatite) and silica are calculated with a distribution of hollow spheres. 
For the amorphous olivine and pyroxene particles we use Mie theory since for
these species the effects of shape on absorption properties are minor.

The $\kappa_i$ are calculated using laboratory measurements of the
refractive index as a function of wavelength. References for the measurements used
for the various dust species are listed in Table\,\ref{tab:t2_fs:template
data}.

\begin{table*}[!t]
\begin{center}
\begin{tabular}{llll}
\hline
Dust component & Chemical formula & Shape & Reference\\
\hline
Amorphous olivine	&Mg$_{2x}$Fe$_{2-2x}$SiO$_4$  &Homogeneous Spheres & Dorschner et al.\citeyear{Dorschner1995} \\
Amorphous pyroxene	&Mg$_{x}$Fe$_{1-x}$SiO$_3$    &Homogeneous Spheres & Dorschner et al.\citeyear{Dorschner1995} \\
Crystalline forsterite	&Mg$_{2}$SiO$_4$  	      &Irregular (DHS)	   & Servoin \& Piriou \citeyear{Servoin1973} \\
Crystalline enstatite	&MgSiO$_3$		      &Irregular (DHS)	   & J\"ager et al. \citeyear{Jaeger1998} \\
Amorphous silica 	&SiO$_2$ 		      &Irregular (DHS)	   & Spitzer \& Kleinman \citeyear{1960PhRv..121.1324S} \\
\hline   
\end{tabular}
\end{center}
\caption{Characteristics of the various dust components used in the fitting procedure
(see section~\ref{sec:t2_fs:compositional_fits}). The chemical formulae and
assumed grain shapes have been indicated in the second and third column. In the fourth
column, we give references for the refractive index data used.}
\label{tab:t2_fs:template data}
\end{table*}

\subsubsection{Size of the dust grains}

The dust grains in circumstellar disks most likely have a rather broad size
distribution. In the 10 micron region the observational data are sensitive to
the dust grains with a volume equivalent radius up to a few micron in size.
Larger grains mainly contribute to the continuum. In order to minimize the
number of free parameters in the fitting procedure, we want to sample the size
distribution carefully. We find that the variety of spectral shapes can be best
covered using only two distinct particle sizes, a `small' particle size with a
material volume equivalent sphere radius $r_V=0.1\,\mu$m and a `large' particle
size with $r_V=1.5\,\mu$m \citep[for a similar approach
see][]{2001A&A...375..950B, 2004ApJ...610L..49H}. We have extensively checked
that the results of the analysis using three, four or five particle sizes with
volume equivalent radii ranging up to $3.5\,\mu$m do not change significantly.
The size of the large grain component, when using only two particle sizes, has
to be chosen with care. When it is too large, the difference in absolute value
of the mass absorption coefficients of the small and the large grain component
will be too big. If, for example, one adopts for the large grain size
$r_V=2\,\mu$m, one would overestimate the abundance of large grains. This is
especially so for the large crystalline grains and results in an overestimate
of the crystallinity. 
The differences between the absolute values of the emissivities of a
0.1 and a 2$\,\mu$m amorphous grain are not so big.
Actually, adopting 2$\,\mu$m sized grains
is even slightly better for the large amorphous component 
\citep[see also][]{2001A&A...375..950B}.
However, considering both the amorphous
and crystalline component, $1.5\,\mu$m is the best choice for the
large grains.

\subsubsection{Fitting procedure and error analysis}

\begin{figure}[!t]
\centerline{
\hspace{-0.15cm}
\includegraphics[width=9.4cm]{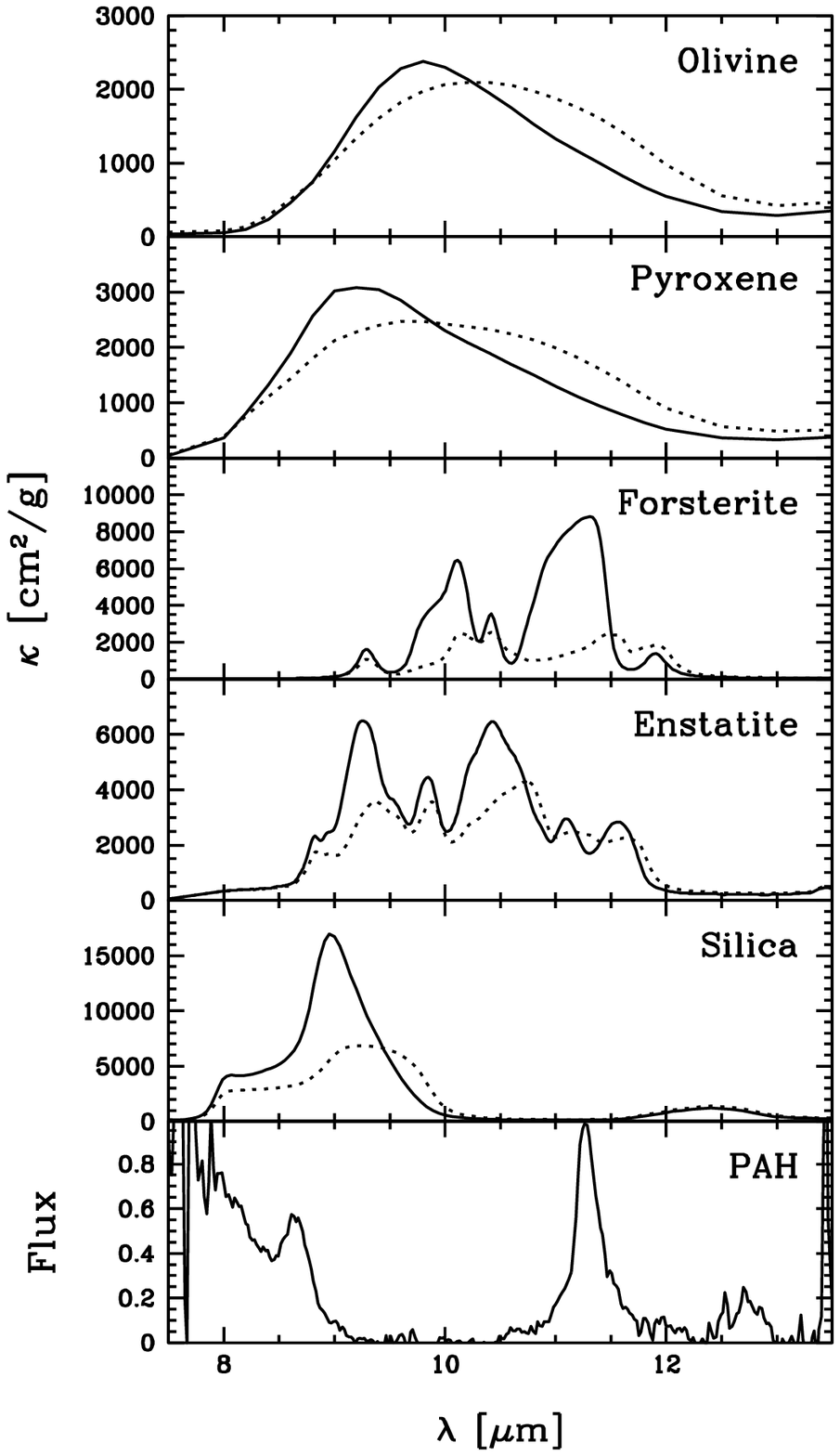}}
\caption{The mass absorption coefficients of the various templates used 
in the fitting procedure (upper 5 panels).
We use grains with volume equivalent radii of 0.1\,$\mu$m (solid lines) and
1.5\,$\mu$m (dotted lines). In the lower panel we show the template
used for the PAH emission, which is normalized such that the maximum flux
in the $8$ to $13$ micron region equals unity. For a detailed discussion see 
section~\ref{sec:t2_fs:compositional_fits}.}
\label{fig:t2_fs:templates}
\end{figure}

In order to keep the number of free parameters in the model small, only two
grain sizes are used for every silicate dust type (as discussed above). 
We thus have contributions of five silicate species, of PAHs and of a blackbody
continuum of which the absolute level and shape (characteristic temperature)
can be varied. The emissivities of the silicates are multiplied by a blackbody
spectrum with the same characteristic temperature as the continuum. This
results in 13 free parameters. The silicate and PAH templates are shown in
Fig.~\ref{fig:t2_fs:templates}.

To fit the spectra we minimize the reduced $\chi^2$ of the entire 10
micron region given by
\begin{equation}
\chi^2=\frac{1}{N_\lambda-M}\sum_{i=1}^{N_\lambda} \left|\frac{\mathcal{F}_\nu^\mathrm{model}(\lambda_i)-\mathcal{F}_\nu^\mathrm{observed}(\lambda_i)}{\sigma_i}\right|^2.
\end{equation}
Here $N_\lambda$ is the number of wavelength points $\lambda_i$, $M$
is the number of fit parameters (in this case $M=13$) and $\sigma_i$
is the absolute error on the observed flux at wavelength
$\lambda_i$. For a given characteristic temperature $T_c$ we can calculate the optimal
values for the weights $w_i$ of the individual dust components,
using a linear least square fitting
procedure.

The measurement errors ($\sigma_i$) used in the fitting procedure
represent the statistical noise in the spectra. The calibrator spectra
all have a very high signal to noise ratio (SNR)
and statistical noise of the calibration
observations is negligible. The SNR in our Herbig star spectra range from
$\sim$18 in the faintest source (HD\,135344) to approximately 60 in
the bright sources (e.g. HD\,100546).
There are also systematic uncertainties,
arising from an imperfect calibration, and the uncertainty in the used
spectral templates for the calibrators. Some degree of systematic
error is inevitable, since the science target and calibrator cannot be
measured at the same time and in the same direction. Since we cannot
assess the systematic uncertainties we do not take these into
account. We note, however, that agreement between our ground based
spectra and high SNR ISO spectra is generally very good. For the faint
sources (such as UX~Ori), the statistical noise dominates the error
budget. For bright sources (e.g. HD\,144432) the systematic
uncertainties may be important, implying that we underestimate
the errors. This will evidently lead to higher $\chi^2$ values in the
fit procedure.

The errors on the fit parameters are calculated using a Monte Carlo
method. For every spectrum we generate 1000 synthetic spectra,
by randomly adding Gaussian noise to the spectrum with
a distribution of width $\sigma_i$ at each wavelength point. 
This yields 1000 spectra that are all consistent with our
data. On each of these, we perform the exact same compositional fit
procedure, yielding (slightly) different values of the fit
parameters.  From the resulting distribution of all fit parameters, we
calculate the mean (which will be our 'best fit' value) and standard
deviation. Besides its simplicity, this method has the advantage that
degeneracies between fit parameters automatically show up as large
errors in these parameters.

\begin{figure}[!t]
\centerline{
  \includegraphics[width=9.1cm]{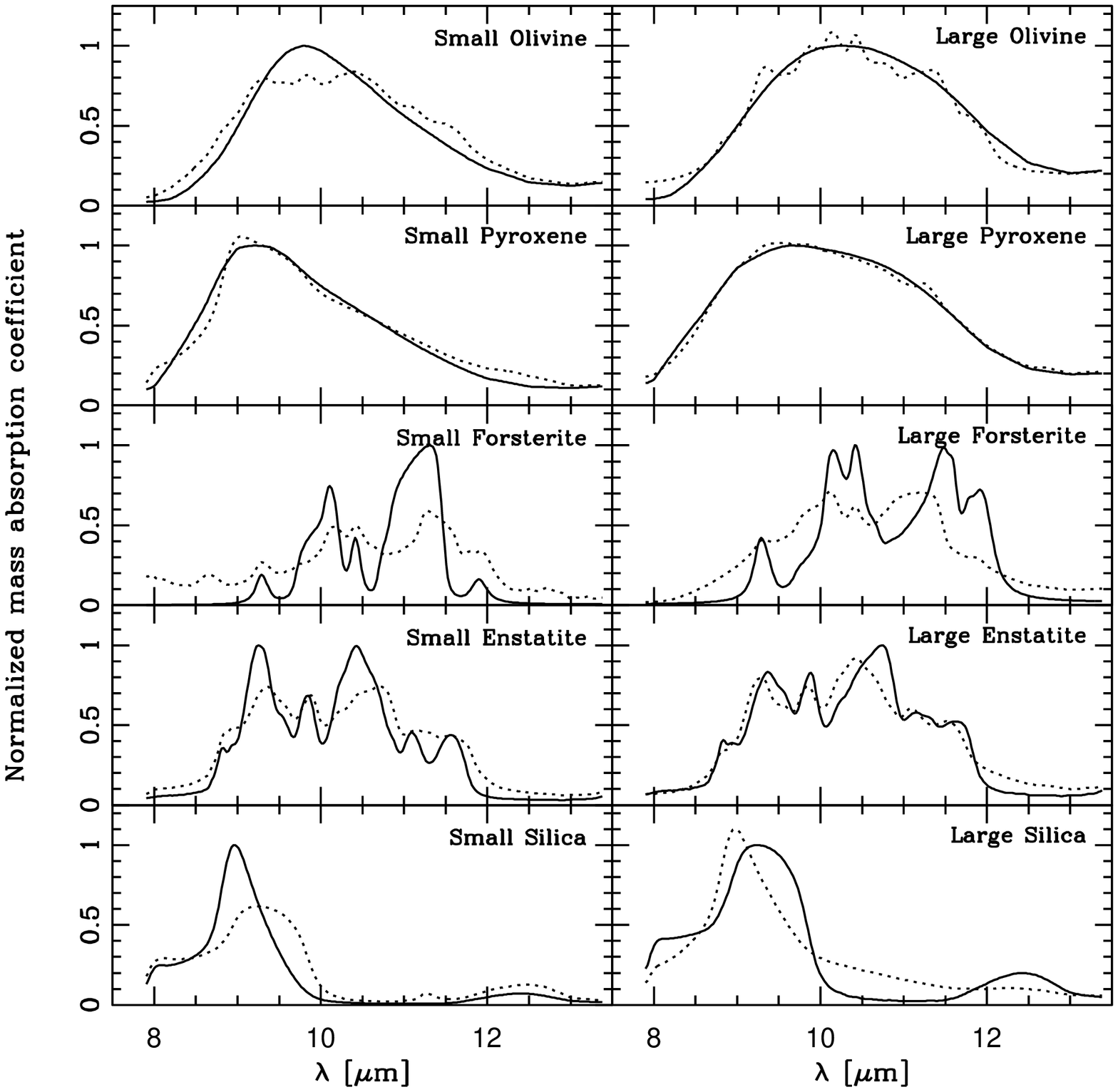}}
\caption{The templates used in the fitting procedure (solid curves)
together with the resulting best fits using all other templates
(dotted curves).  The spectra are all normalized such that the maximum
value of the template equals unity. In the left column we show the
templates for the small grains (0.1\,$\mu$m), in the right column
those of the large grains (1.5\,$\mu$m). 
}
\label{fig:t2_fs:degeneracies} 
\end{figure}

\subsection{Results}
\label{sec:t2_fs:results}

Fig.\,\ref{fig:t2_fs:all_spectra} shows the observed spectra together
with the best fit model spectra. For comparison we also fit the
interstellar extinction as observed towards the galactic center
\citep{2004ApJ...609..826K}.
The resulting values of the fit
parameters are summarized in Table\,\ref{tab:t2_fs:fits}. The overall
quality of the model fits is very good and we thus conclude that the
diversity of the shapes of the observed spectra are well covered by
the choice of the spectral templates. We notice that in some sources
the model fits show slightly more spectral structure between 9 and
11\,$\mu$m than the observed spectra.  This could be caused by the
choice of the shape distribution of the dust grains, which may be too
simple to represent the spectral details of realistic particles. We
also notice that in, for example, HD\,179218 we cannot accurately
reproduce the detailed shape of the spectrum. Especially the feature
around 11.3\,$\mu$m is less sharply peaked in the observed spectrum
than in the model fit. This effect can also be seen in other sources,
albeit in a more modest form, and could be caused by a missing dust
component or by the choice of the shape distribution.

\hspace{-4cm}

\begin{sidewaystable*}

\begin{center}

\caption{The best fit values of the parameters in our compositional fits.
The abundances of small (0.1\,$\mu$m) and large (1.5\,$\mu$m) grains of the
various dust species are given
as fractions of the total dust mass, \emph{excluding} the dust responsible
for the continuum emission. If a species was not found, or unconstrained
by the spectra, this is indicated by a -~symbol.
The PAH and continuum flux contributions
(the last two columns) are listed as percentages of the total integrated flux over the 10\,$\mu$m region, contained in these
components. These are measures for the relative flux contributions, but cannot be interpreted as relative dust masses.}

\label{tab:t2_fs:fits}
\linespread{1.3}%
\selectfont
\begin{tabular}{r l l l ll ll ll ll ll l l l}
\hline
\# & Star & $\chi^2$ & $T_c$\,[K] & 
 \multicolumn{2}{c}{Olivine} &
 \multicolumn{2}{c}{Pyroxene} &
 \multicolumn{2}{c}{Forsterite} &
 \multicolumn{2}{c}{Enstatite} &
 \multicolumn{2}{c}{Silica} & PAH & Cont. \\
    &                         &    &                       & Small                   &                   Large &                   Small &                   Large &          Small &                   Large &                   Small &                   Large &                   Small &                   Large &             contr.       &             contr.      \\
\hline
  0 & Gal. Center             &66.8&  -                    & 85.8$_{-  0.1}^{+  0.1}$&  -                      & 12.5$_{-  0.1}^{+  0.1}$&  -                      &   0.6$_{-  0.0}^{+  0.0}$&  -                      &  -                      &  -                      &  -                      &  1.0$_{-  0.0}^{+  0.0}$&  0.5$_{-  0.0}^{+  0.0}$& -                       \\
  1 & AB Aur                  & 2.5&464$_{- 4}^{+ 6}$& 49.5$_{-  2.4}^{+  2.2}$& 47.2$_{-  2.6}^{+  2.5}$&  -                      &  -                      &   0.6$_{-  0.2}^{+  0.2}$&  2.7$_{-  0.7}^{+  0.7}$&  -                      &  -                      &  -                      &  -                      &  1.9$_{-  0.2}^{+  0.1}$& 53.1$_{-  0.4}^{+  0.5}$\\
  2 & UX Ori                  & 0.7&677$_{-38}^{+29}$& 63.6$_{-  4.5}^{+  4.9}$& 30.7$_{-  7.9}^{+  6.0}$&  -                      &  2.0$_{-  1.8}^{+  5.8}$&   1.2$_{-  0.4}^{+  0.4}$&  0.3$_{-  0.3}^{+  0.9}$&  0.6$_{-  0.4}^{+  0.6}$&  1.1$_{-  0.8}^{+  1.2}$&  -                      &  0.6$_{-  0.4}^{+  0.4}$&  0.1$_{-  0.1}^{+  0.3}$& 45.2$_{-  1.4}^{+  1.3}$\\
  3 & HD\,36112                & 1.7&669$_{-21}^{+19}$& 50.5$_{-  1.7}^{+  1.8}$& 22.4$_{-  3.1}^{+  3.2}$&  -                      & 22.2$_{-  1.8}^{+  1.8}$&   2.9$_{-  0.1}^{+  0.1}$&  0.0$_{-  0.0}^{+  0.1}$&  1.8$_{-  0.2}^{+  0.2}$&  0.2$_{-  0.2}^{+  0.4}$&  -                      &  -                      &  0.1$_{-  0.1}^{+  0.1}$& 33.9$_{-  0.5}^{+  0.4}$\\
  4 & HK Ori                  & 0.4&516$_{-11}^{+12}$&  7.1$_{-  5.4}^{+  7.6}$&  0.9$_{-  0.9}^{+ 16.3}$& 13.7$_{-  7.9}^{+  7.2}$& 75.3$_{- 13.3}^{+  9.9}$&   0.7$_{-  0.6}^{+  1.0}$&  1.4$_{-  1.3}^{+  2.7}$&  0.1$_{-  0.1}^{+  1.0}$&  0.1$_{-  0.1}^{+  1.6}$&  -                      &  0.8$_{-  0.8}^{+  1.6}$&  2.5$_{-  0.4}^{+  0.4}$& 68.5$_{-  1.3}^{+  1.4}$\\
  5 & HD\,245185               & 0.6&337$_{- 9}^{+ 7}$& 28.5$_{-  3.5}^{+  3.5}$& 57.4$_{-  8.8}^{+  7.5}$&  0.1$_{-  0.1}^{+  2.5}$& 12.2$_{-  5.9}^{+  6.5}$&   0.2$_{-  0.2}^{+  0.3}$&  0.7$_{-  0.6}^{+  0.9}$&  0.0$_{-  0.0}^{+  0.3}$&  0.0$_{-  0.0}^{+  0.7}$&  -                      &  0.9$_{-  0.5}^{+  0.5}$&  1.1$_{-  0.3}^{+  0.2}$& 50.6$_{-  1.1}^{+  1.1}$\\
  6 &  V380 Ori               & 1.8&481$_{- 1}^{+ 9}$&  -                      &  0.8$_{-  0.8}^{+  6.5}$&  -                      & 74.6$_{-  5.0}^{+  2.9}$&   2.2$_{-  0.5}^{+  0.6}$& 10.8$_{-  1.4}^{+  1.5}$&  -                      & 10.8$_{-  1.7}^{+  1.8}$&  0.8$_{-  0.3}^{+  0.3}$&  0.0$_{-  0.0}^{+  0.4}$&  0.7$_{-  0.1}^{+  0.1}$& 84.9$_{-  0.4}^{+  0.5}$\\
  7 & HD\,37357                & 0.5&508$_{-25}^{+22}$&  3.9$_{-  3.4}^{+  5.9}$& 39.4$_{- 14.8}^{+ 11.9}$&  0.1$_{-  0.1}^{+  3.6}$& 46.0$_{-  9.9}^{+ 10.4}$&   2.7$_{-  0.6}^{+  0.5}$&  2.2$_{-  1.5}^{+  1.8}$&  0.9$_{-  0.7}^{+  0.9}$&  1.3$_{-  1.1}^{+  1.8}$&  0.0$_{-  0.0}^{+  0.2}$&  3.5$_{-  0.8}^{+  0.8}$&  0.1$_{-  0.1}^{+  0.3}$& 47.5$_{-  1.5}^{+  1.6}$\\
  8 & HD\,37806                & 1.8&525$_{- 6}^{+ 7}$&  -                      & 59.8$_{-  5.7}^{+  4.6}$&  -                      &  0.8$_{-  0.8}^{+  5.2}$&   6.2$_{-  0.8}^{+  0.9}$&  3.5$_{-  1.8}^{+  1.9}$&  0.2$_{-  0.2}^{+  0.9}$& 20.9$_{-  2.6}^{+  2.7}$&  5.0$_{-  0.4}^{+  0.4}$&  3.5$_{-  0.8}^{+  0.8}$&  1.8$_{-  0.2}^{+  0.2}$& 73.7$_{-  0.9}^{+  1.0}$\\
  9 & HD\,95881                & 3.7&429$_{- 9}^{+ 2}$&  -                      &  -                      &  -                      & 79.7$_{-  2.7}^{+  4.0}$&   0.7$_{-  0.5}^{+  0.5}$&  4.7$_{-  1.9}^{+  1.6}$&  0.0$_{-  0.0}^{+  0.8}$& 11.3$_{-  2.4}^{+  1.7}$&  3.0$_{-  0.3}^{+  0.3}$&  0.5$_{-  0.5}^{+  0.9}$&  3.0$_{-  0.1}^{+  0.2}$& 84.2$_{-  0.7}^{+  0.4}$\\
 10 & HD\,97048                & 1.4&303$_{- 3}^{+ 7}$&  -                      &  -                      &  -                      &  -                      &   -                      &  -                      &  -                      &  -                      &  -                      &  -                      & 14.1$_{-  0.2}^{+  0.2}$& 83.1$_{-  0.4}^{+  0.6}$\\
 11 & HD\,100453               & 2.1&429$_{- 9}^{+ 1}$&  -                      &  -                      &  -                      &  -                      &   -                      &  -                      &  -                      &  -                      &  -                      &  -                      &  4.4$_{-  0.1}^{+  0.1}$& 93.5$_{-  0.4}^{+  0.2}$\\
 12 & HD\,100546               & 5.6&261$_{- 1}^{+ 9}$&  2.5$_{-  1.6}^{+  1.1}$& 78.5$_{-  1.2}^{+  1.3}$&  -                      &  0.0$_{-  0.0}^{+  0.7}$&   6.2$_{-  0.2}^{+  0.8}$&  0.1$_{-  0.1}^{+  0.4}$&  -                      &  5.7$_{-  0.4}^{+  0.5}$&  0.2$_{-  0.1}^{+  0.1}$&  6.7$_{-  0.6}^{+  0.3}$&  3.7$_{-  0.7}^{+  0.1}$& 55.3$_{-  0.6}^{+  2.9}$\\
 13 & HD\,101412               & 0.5&425$_{- 5}^{+ 6}$&  -                      &  0.3$_{-  0.3}^{+  9.1}$&  0.2$_{-  0.2}^{+  4.3}$& 75.2$_{-  7.0}^{+  5.4}$&   0.7$_{-  0.6}^{+  1.2}$&  3.4$_{-  2.5}^{+  3.1}$&  0.1$_{-  0.1}^{+  1.2}$& 16.8$_{-  3.4}^{+  3.9}$&  3.2$_{-  0.6}^{+  0.6}$&  0.1$_{-  0.1}^{+  1.2}$&  2.0$_{-  0.2}^{+  0.2}$& 80.2$_{-  0.9}^{+  1.1}$\\
 14 & HD\,104237               & 3.9&446$_{- 6}^{+ 4}$&  -                      &  -                      &  -                      & 77.7$_{-  1.3}^{+  1.6}$&   5.9$_{-  0.6}^{+  0.4}$&  -                      &  -                      & 11.8$_{-  1.2}^{+  1.0}$&  1.7$_{-  0.1}^{+  0.2}$&  2.9$_{-  0.4}^{+  0.4}$&  1.4$_{-  0.1}^{+  0.2}$& 63.4$_{-  0.9}^{+  0.6}$\\
 15 & HD\,135344               & 0.8&680$_{-14}^{+18}$&  -                      &  -                      &  -                      &  -                      &   -                      &  -                      &  -                      &  -                      &  -                      &  -                      &  4.9$_{-  0.3}^{+  0.3}$& 90.3$_{-  1.0}^{+  1.0}$\\
 16 & HD\,139614               & 0.6&277$_{- 7}^{+ 4}$& 16.0$_{-  6.5}^{+  6.2}$& 47.0$_{- 16.7}^{+ 11.2}$&  -                      & 29.7$_{-  9.8}^{+ 13.8}$&   2.4$_{-  0.8}^{+  0.9}$&  0.7$_{-  0.6}^{+  1.7}$&  0.0$_{-  0.0}^{+  0.7}$&  4.1$_{-  2.1}^{+  2.3}$&  -                      &  0.1$_{-  0.1}^{+  0.7}$&  2.2$_{-  0.3}^{+  0.3}$& 75.9$_{-  1.6}^{+  1.3}$\\
 17 & HD\,142666               & 1.6&396$_{- 6}^{+ 4}$& 42.8$_{-  2.7}^{+  2.9}$&  9.7$_{-  6.5}^{+  5.0}$&  -                      & 43.5$_{-  2.9}^{+  3.9}$&   1.1$_{-  0.3}^{+  0.3}$&  2.3$_{-  0.8}^{+  0.8}$&  0.0$_{-  0.0}^{+  0.3}$&  0.6$_{-  0.5}^{+  0.8}$&  -                      &  -                      &  1.7$_{-  0.1}^{+  0.2}$& 66.8$_{-  0.5}^{+  0.5}$\\
 18 & HD\,142527               &11.1&523$_{- 3}^{+ 8}$&  -                      &  0.0$_{-  0.0}^{+  0.8}$&  -                      & 73.2$_{-  1.0}^{+  0.7}$&   8.6$_{-  0.2}^{+  0.4}$&  -                      &  0.6$_{-  0.2}^{+  0.2}$& 14.2$_{-  0.6}^{+  0.8}$&  2.4$_{-  0.1}^{+  0.1}$&  1.0$_{-  0.3}^{+  0.2}$&  3.1$_{-  0.2}^{+  0.1}$& 58.8$_{-  0.3}^{+  0.7}$\\
 19 & HD\,144432               & 3.0&401$_{- 1}^{+10}$& 52.7$_{-  0.7}^{+  0.7}$&  0.0$_{-  0.0}^{+  1.1}$&  -                      & 42.3$_{-  0.8}^{+  0.7}$&   1.9$_{-  0.1}^{+  0.1}$&  0.7$_{-  0.3}^{+  0.3}$&  0.7$_{-  0.2}^{+  0.2}$&  0.9$_{-  0.3}^{+  0.3}$&  -                      &  0.6$_{-  0.1}^{+  0.1}$&  0.5$_{-  0.1}^{+  0.1}$& 39.7$_{-  0.3}^{+  0.5}$\\
 20 & HD\,144668               & 9.6&517$_{- 7}^{+ 3}$&  -                      & 56.9$_{- 13.2}^{+  6.8}$&  -                      & 12.0$_{-  6.1}^{+ 12.4}$&   7.0$_{-  0.3}^{+  0.3}$&  6.3$_{-  0.8}^{+  0.9}$&  0.0$_{-  0.0}^{+  0.3}$& 12.1$_{-  0.8}^{+  0.9}$&  1.5$_{-  0.3}^{+  0.2}$&  4.2$_{-  0.3}^{+  0.3}$&  1.2$_{-  0.1}^{+  0.1}$& 85.3$_{-  0.3}^{+  0.2}$\\
 21 & HD\,150193               & 5.1&410$_{-10}^{+ 0}$&  -                      &  0.7$_{-  0.6}^{+  1.1}$&  -                      & 82.1$_{-  0.8}^{+  0.7}$&   3.3$_{-  0.1}^{+  0.1}$&  4.1$_{-  0.3}^{+  0.3}$&  0.2$_{-  0.2}^{+  0.2}$&  3.7$_{-  0.4}^{+  0.3}$&  -                      &  5.8$_{-  0.1}^{+  0.1}$&  0.2$_{-  0.1}^{+  0.1}$& 39.7$_{-  0.3}^{+  0.2}$\\
 22 & HD\,163296               &10.7&461$_{- 1}^{+ 9}$& 15.3$_{-  1.4}^{+  0.7}$& 29.8$_{-  1.4}^{+  3.8}$&  -                      & 42.2$_{-  2.1}^{+  0.8}$&   3.1$_{-  0.1}^{+  0.1}$&  1.3$_{-  0.3}^{+  0.3}$&  0.2$_{-  0.1}^{+  0.1}$&  4.0$_{-  0.3}^{+  0.3}$&  -                      &  4.1$_{-  0.1}^{+  0.1}$&  1.1$_{-  0.1}^{+  0.1}$& 49.6$_{-  0.2}^{+  0.2}$\\
 23 & HD\,169142               & 0.9&348$_{- 8}^{+ 4}$&  -                      &  -                      &  -                      &  -                      &   -                      &  -                      &  -                      &  -                      &  -                      &  -                      & 16.1$_{-  0.2}^{+  0.2}$& 82.0$_{-  0.5}^{+  0.4}$\\
 24 & HD\,179218               &11.8&350$_{-14}^{+ 0}$&  -                      & 11.3$_{-  2.6}^{+  2.6}$&  -                      & 58.7$_{-  2.0}^{+  2.1}$&   -                      &  3.5$_{-  0.7}^{+  0.6}$&  6.2$_{-  0.4}^{+  0.4}$& 18.3$_{-  0.7}^{+  0.8}$&  2.0$_{-  0.1}^{+  0.1}$&  -                      &  4.4$_{-  0.1}^{+  0.1}$& 64.8$_{-  0.2}^{+  0.2}$\\
\hline
\end{tabular}
\linespread{1}%
\selectfont
\end{center}
\end{sidewaystable*}

As an objective measure for the goodness of fit, the reduced $\chi^2$
of every fit is listed in Table\,\ref{tab:t2_fs:fits}. For a good fit
this parameter should be close to unity.  The likely reason that we
have relatively high values of $\chi^2$ for about half of the sources
is that we do not take into account the uncertainties on the
$\kappa_i$.  These are mainly caused by uncertainties in
the shape, structure and size of the grains, and in the laboratory
measurements of the wavelength dependent refractive indices.

To test whether we have degeneracies between the various templates used in the
fitting procedure we tried to fit each of the silicate templates using
a linear combination of all other templates. The results are shown
in Fig.~\ref{fig:t2_fs:degeneracies} where we plot the mass
absorption coefficients together with the best fit using the other
templates (here the mass absorption coefficients have been normalized such
that the maximum value equals unity).
The figure shows that almost all of the templates used
have a unique spectral structure that cannot be reproduced by the
other templates. 
Only the emission
from large pyroxene grains can be reproduced reasonably well by the
other templates, although significant differences still exist (e.g.
the 11.3\,$\mu$m feature that is present in the fit to the large
pyroxene opacity in Fig.~\ref{fig:t2_fs:degeneracies}).
The fit consists of 52\% large olivine
grains, 43\% small pyroxene grains and only 5\% of crystalline
silicates. This could result in a slight change in the mass
fraction of large grains when this template is not used. The
crystalline fraction would not be affected significantly. Because
the small pyroxene grains are needed in order to reproduce the short
wavelength side of some of the spectra, we chose to include also the
large pyroxene grains for consistency. 
The presence of
large pyroxene grains can be firmly established only in the highest
SNR spectra.

There are a few points one has to keep in mind when interpreting the
results of the analysis presented above.
\begin{itemize} 
\item Since we
consider only the 10~micron region of the infrared spectrum our data are sensitive
only to grains with a temperature of $\sim 200$~K or more. This limits
our study to the inner disk regions (i.e. $\lesssim 10$-20\,AU).  
\item Because of the same limitation of
the spectral range our data are only sensitive to relatively modest sized grains.
In order for the emission to show an observable spectral
signature, the dust grains have to be small compared to the wavelength of
radiation. 10\,$\mu$m measurements are sensitive to grains with a volume
equivalent radius $r_V\lesssim$3\,$\mu$m.  
\item The disk regions that we study are partially optically
thick at 10\,$\mu$m. This means that we cannot see deep into the
disk. We only observe the surface layer.  
\item Due to the limited spatial resolution of our observations, we
observe the integrated spectra of the entire inner disk surface.  The
observed flux is therefore an average over distance to the star and
thus over temperature.  Also, we know from spatially resolved
observations of the innermost part of a few disks that the mineralogy
is not constant as a function of distance to the star 
\citep{2004Natur.432..479V}.
Close to the star we have higher temperatures and
densities which trigger both crystallization and grain growth. 
The derived parameters therefore represent
an average, characteristic temperature and an average dust composition.

\end{itemize}

\subsection{Observed trends in the fits}
\label{sec:t2_fs:observed_trends}

\begin{figure*}[t]
  \hspace{-0.5cm}
  \centerline{
\resizebox{\hsize}{!}{\includegraphics{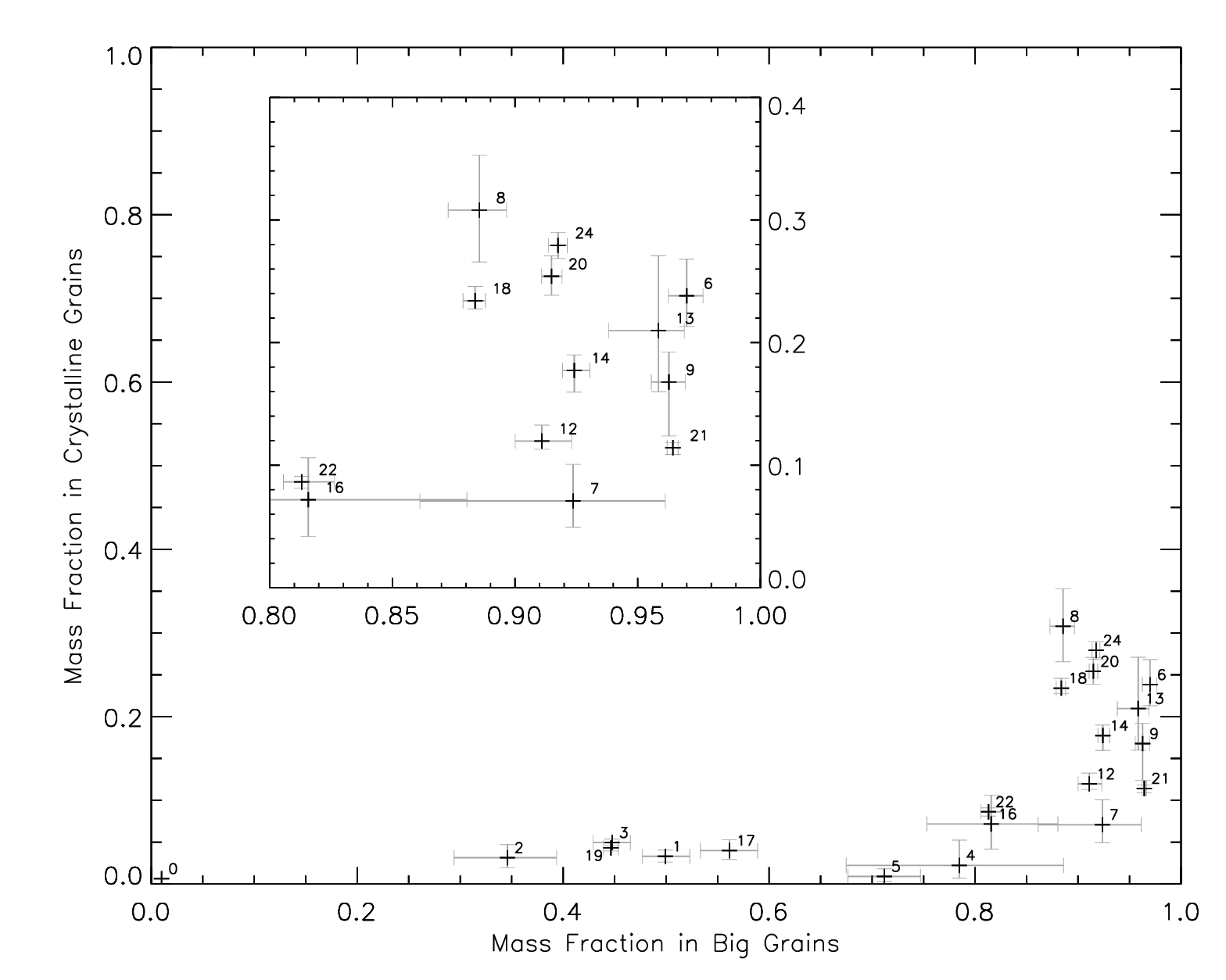}}}

  \caption{ The relation between grain size and crystallinity found in
our spectral fits.  Vertically, we plot the mass fraction of
crystalline grains (forsterite and enstatite).  Horizontally,
the mass fraction of large (1.5\,$\mu$m) grains is plotted.  Since
group\,Ib sources do not display a silicate feature, they are not shown
here.  For a discussion see sections~\ref{sec:t2_fs:observed_trends}
and~\ref{sec:t2_fs:discussion}. }
  \label{fig:t2_fs:Xsil_vs_growth}
\end{figure*}

We will now discuss the trends and correlations observed in the 
derived fit parameters.

In Fig.\,\ref{fig:t2_fs:Xsil_vs_growth} we visualize the grain growth
and crystallinity, as implied by our compositional fits.
Horizontally we plot the mass in large (1.5\,$\mu$m) grains, as a
fraction of the total dust mass, \emph{excluding the dust responsible
for the continuum component, and the PAHs}.  Vertically, we
likewise plot the mass fraction contained in crystalline silicates
in small and large grains,
which is also referred to as the \emph{crystallinity} of the material.

Upon inspection of the figure it is clear
that all disks show signs of substantial removal of small
grains. There are \emph{no} sources with a mass fraction 
in large grains below 30\%.
This infers that none of the sources in our sample contains truly
``pristine'' dust. All sources have an appreciable amount of large
grains at their disk surface, compared to ISM conditions.  In addition,
all sources have a crystallinity that is higher 
than the value we derive for the ISM 
($\sim$0.6\%; Kemper et al. \citeyear{2004ApJ...609..826K}
derive an even more stringent upper limit of 0.4\%).

The derived mass fraction in large grains ranges from $\approx$30\% to
$\approx$100\%, with most sources at high values.
\emph{All sources exhibiting a
high degree of crystallinity have a high mass fraction in large
grains}. There are \emph{no} highly crystalline sources (crystallinity
above 10\,\%) with less than 85\% of the dust mass in large grains.

There are no sources with a mass fraction of crystalline
material above 35\% 
(see also Table\,\ref{tab:t2_fs:fits}).
It should be kept in mind that the silicate emission we see likely
originates in the surface layer of the disk.
Van Boekel et al. (\citeyear{2003A&A...400L..21V}) argue that the disks are
well mixed in the vertical direction, and that therefore the observed
silicate emission should be representative of the whole micron and sub-micron
sized dust population of the disk.  We point out that in this work, ``crystallization''
refers to the process of crystallizing the material (by whatever
means), spreading it over a significant part of the disk region seen
at 10\,$\mu$m, \emph{and} bringing it up to the disk surface where we can see
it spectroscopically. In a scenario where the crystalline silicates
are produced by thermal annealing in the innermost disk regions, and
transported outward by radial mixing, the degree of crystallization is
therefore a measure of the degree of mixing in the disk rather than
the actual process of annealing, which is effectively instantaneous at
the inner disk edge.

\begin{figure}[t]
  \centerline{
  \hspace{-0.4cm}
  \includegraphics[width=9.1cm]{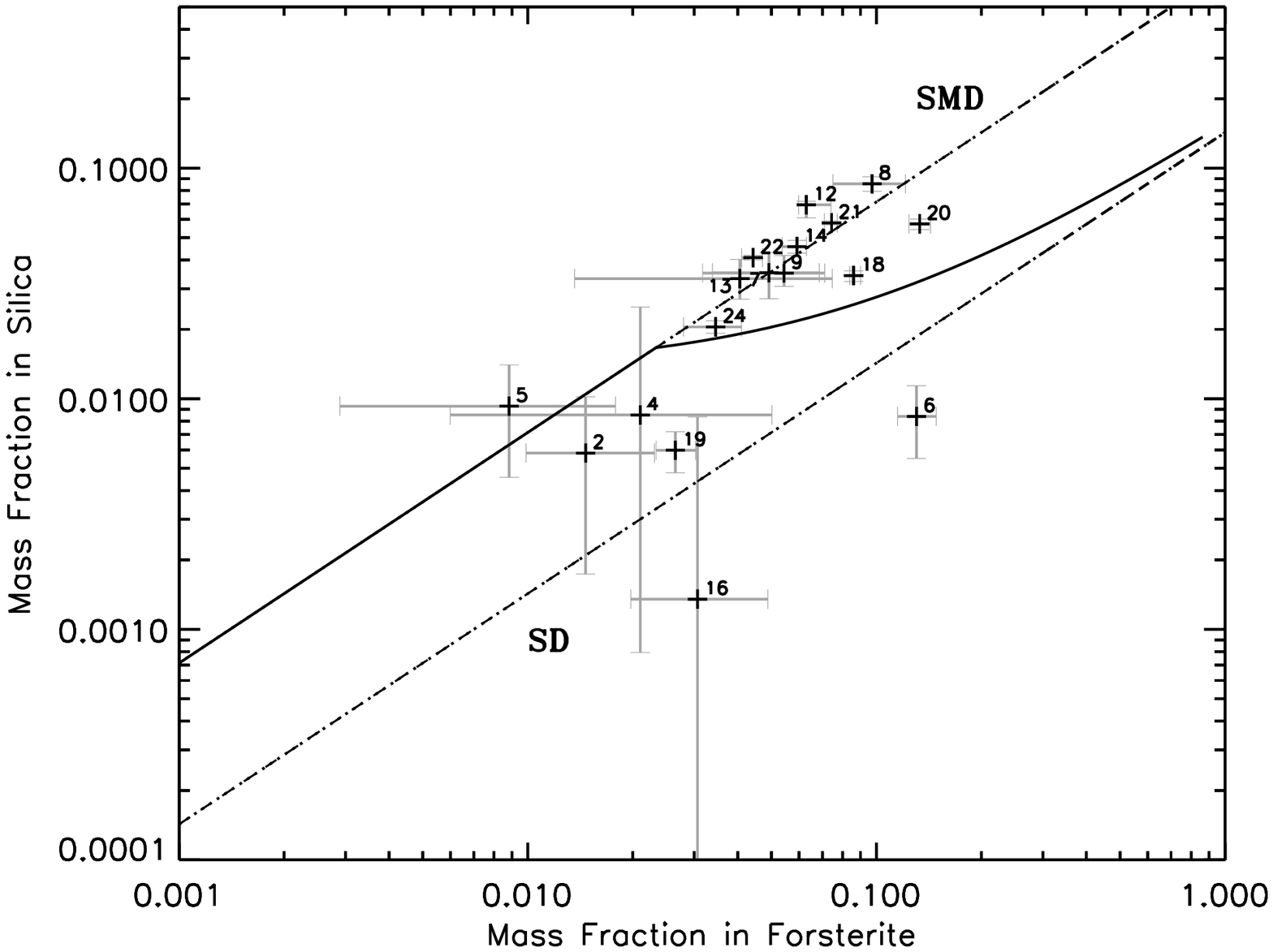}}
  \caption{The fraction of the dust mass contained in silica grains vs. 
the mass fraction contained in forsterite grains. Following 
\cite{2001A&A...375..950B}
we also plot the theoretical annealing behavior of two
different amorphous magnesium silicates, smectite dehydroxylate
(SMD; Mg$_6$Si$_8$O$_{22}$; upper dashed line)
and serpentine dehydroxylate
(SD; Mg$_3$Si$_2$O$_7$; lower dashed line).
The solid line represents
the expected annealing behavior of a mixture of these two silicates,
consisting of
4\% SMD and 96\% SD, which was found by \cite{2001A&A...375..950B}
to give the best fit to their data. Our data are in better agreement
with a pure SMD initial composition.}
  \label{fig:t2_fs:silica_vs_forst}
\end{figure}

In Fig.~\ref{fig:t2_fs:silica_vs_forst} we plot the mass fraction of
dust contained in forsterite versus that contained in silica
grains. It is clear that these mass fractions are correlated.
Experiments show that when forsterite is created by annealing of an
amorphous silicate, silica is formed as a by-product
\citep{1986Icar...66..211R,1997Ap&SS.255..427H,2000A&A...364..282F}.
The amount of silica that is created when forming a certain amount of
forsterite depends on the type of amorphous silicate one starts out
with. As already suggested by \citet{2001A&A...375..950B}, we can try
to constrain the composition of the amorphous material by measuring
the ratio of forsterite over silica.  In the figure
we also plot the expected
annealing behaviour of smectite dehydroxylate (SMD;
Mg$_6$Si$_8$O$_{22}$) and serpentine dehydroxylate (SD;
Mg$_3$Si$_2$O$_7$). In low temperature condensation experiments, these
are the only magnesium silicates that are formed
\citep{1999ApJ...527..395R}.  Whereas \citet{2001A&A...375..950B}
found that a mixture of 4\% of SMD and 96\% of SD yielded the best fit
to their data, our results seem to favour an initial composition of
pure SMD.  Possibly, the discrepancy between the results found by
\citet{2001A&A...375..950B} and the results found here is connected to
the differences in the dust components used to fit the 10\,$\mu$m
spectra. Notably, \citet{2001A&A...375..950B} do not include large
silica and forsterite particles.

\begin{figure}[t]
  \centerline{
  \hspace{-0.6cm}
  \includegraphics[width=9.4cm]{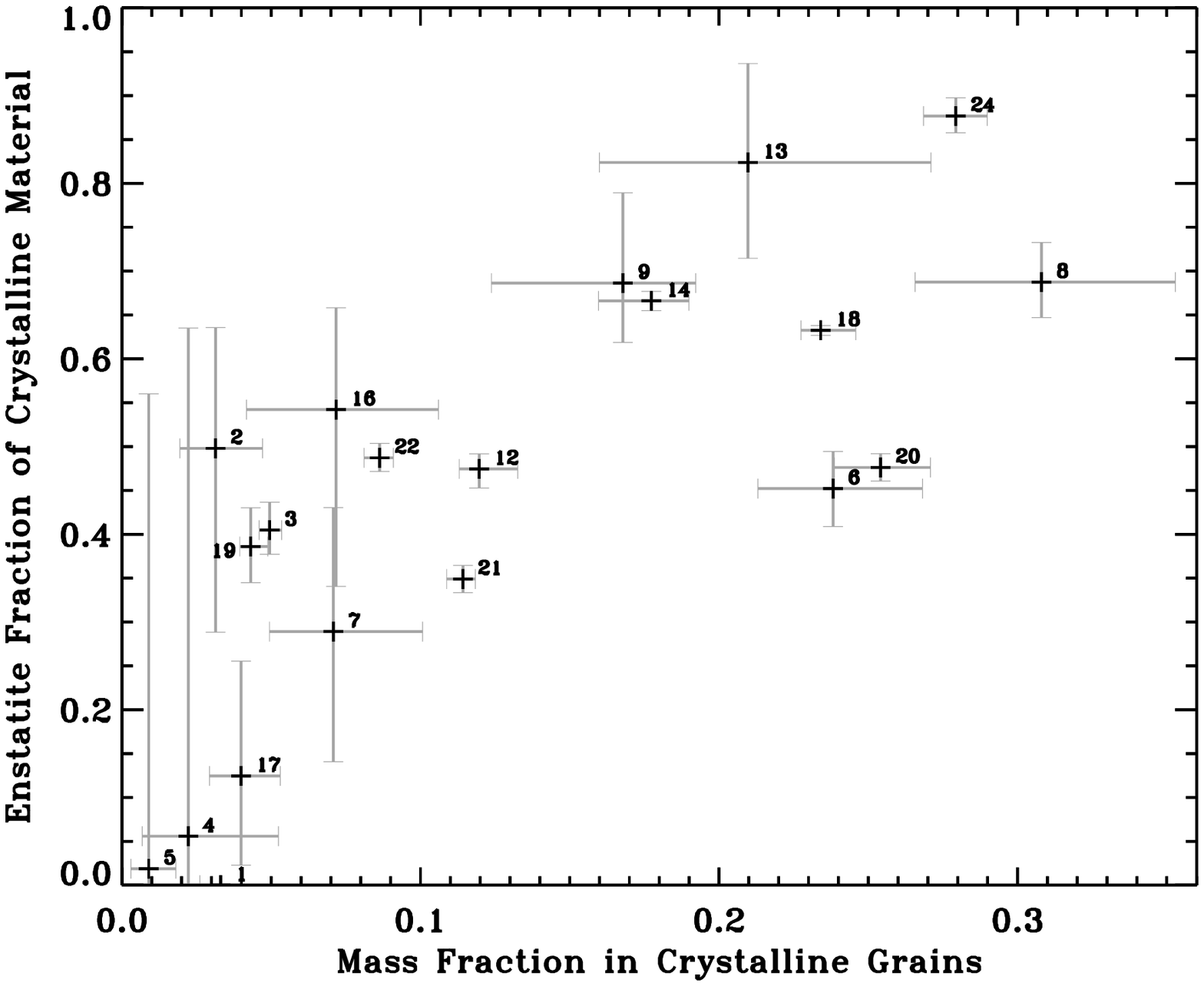}}
  \caption{The mass fraction of crystalline material contained in
enstatite vs. the total mass fraction of crystalline material.  A
value of~0 on the vertical axis indicates that all the crystalline
silicates present in the disk are in the form of forsterite, while a
value of~1 means all crystalline silicates are in the form of
enstatite.}
  \label{fig:t2_fs:forst_over_enst_vs_Xsil}
\end{figure}

In Fig.~\ref{fig:t2_fs:forst_over_enst_vs_Xsil} we show the mass
fraction of crystalline material contained in enstatite versus the
total mass fraction of crystalline material.
There is a correlation between the fraction of the total mass
contained in crystalline silicates and the composition of these
crystalline silicates. In general, for sources with a high degree of
crystallinity most crystals are in the form of enstatite, while for
the sources with a low crystallinity, forsterite is the dominant
crystalline species.
We will discuss this further in section~\ref{sec:t2_fs:new_constraints}.

\begin{figure}[t]
  \centerline{
  \hspace{-0.6cm}
  \includegraphics[width=9.4cm]{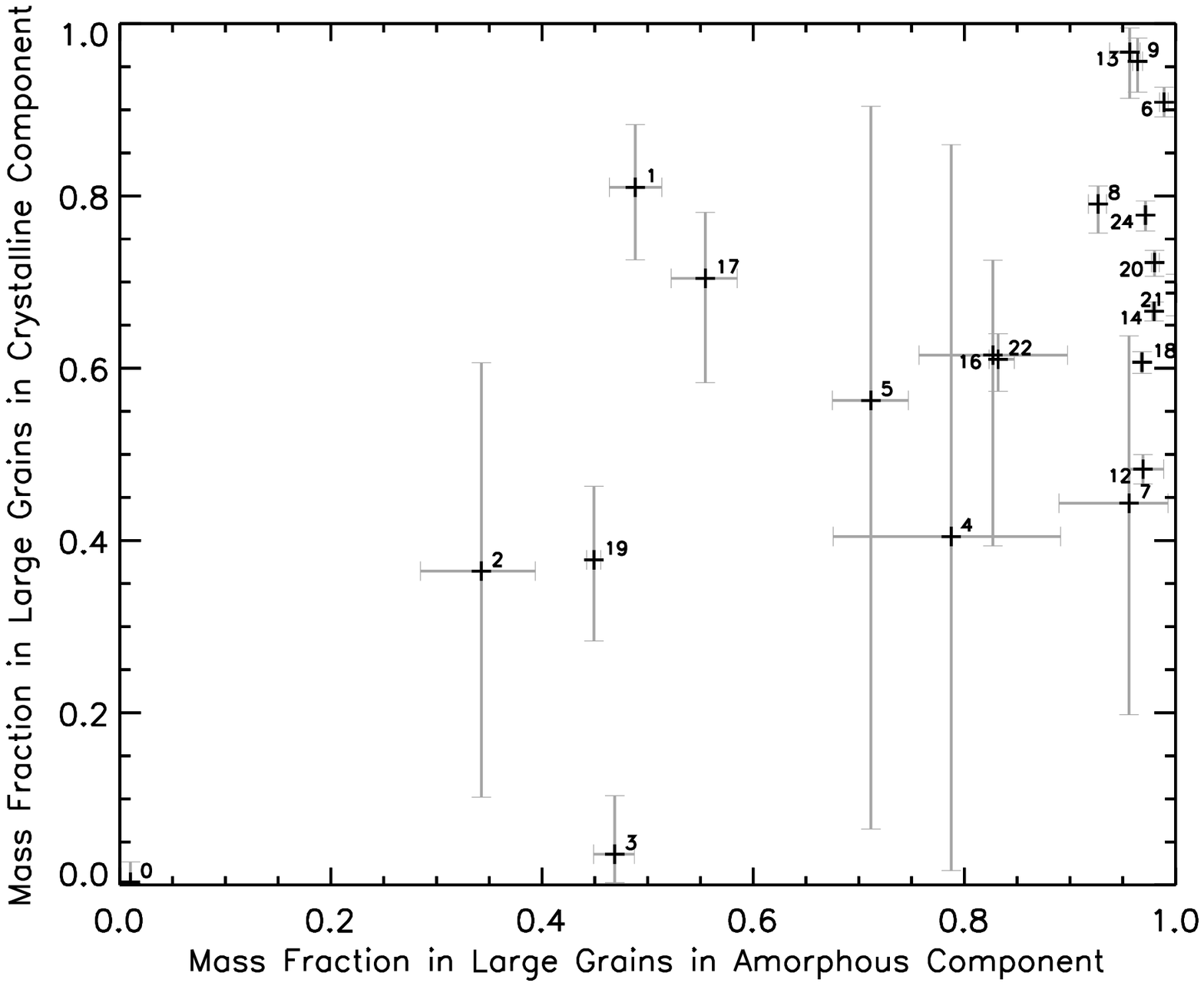}}
  \caption{The mass fraction of large grains in the crystalline
grain population vs. the mass fraction of large grains in the amorphous
component. When the average size of the amorphous grains is large,
the majority of the crystalline material resides in large grains
as well.}
  \label{fig:t2_fs:growth_crys_vs_growth_amorph}
\end{figure}

The amount of growth that the crystalline material has experienced is
compared to the growth in the amorphous component in
Fig.\,\ref{fig:t2_fs:growth_crys_vs_growth_amorph}.  If the amorphous
grains are large, also the crystalline grains are large, though the
correlation is not tight. In all sources in which the amorphous
component has more than 85\% large grains, also the crystalline
component is dominated by large grains. The sources that have less
than 85\% large grains in the amorphous component all have a low
crystallinity (see Fig.\,\ref{fig:t2_fs:Xsil_vs_growth}). Therefore,
the ratio of large and small crystals is poorly constrained in these
sources, which is reflected in the large errorbars.

\begin{figure}[t]
  \centerline{
  \hspace{-0.6cm}
  \includegraphics[width=9.4cm]{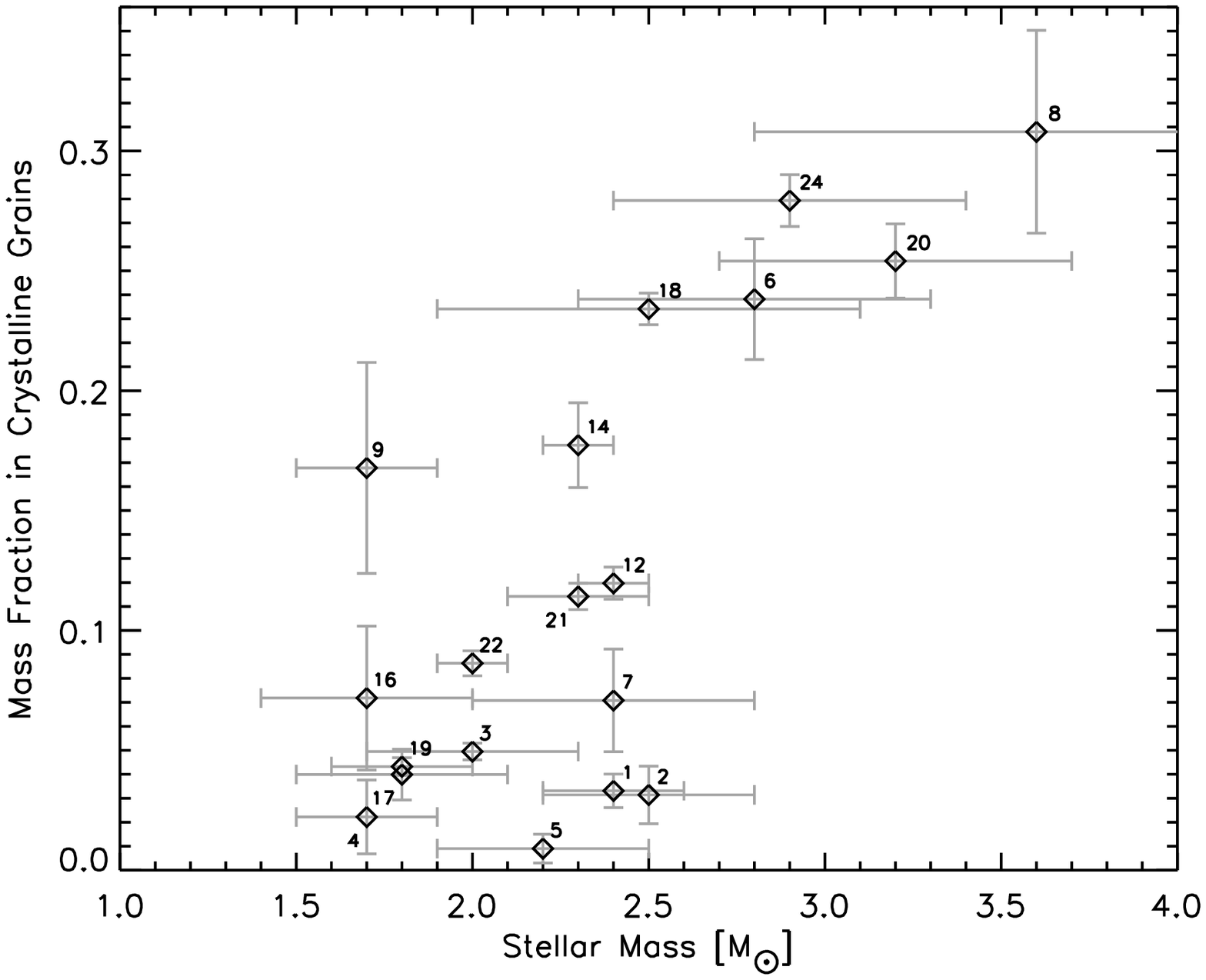}}
  \caption{The mass fraction of dust in crystalline grains vs.
the stellar mass. Higher mass stars show an on average higher fraction
of crystalline grains than do lower mass stars.}
  \label{fig:t2_fs:Xsil_vs_Mstar}
\end{figure}

\begin{figure}[t]
  \centerline{
  \hspace{-0.6cm}
  \includegraphics[width=9.4cm]{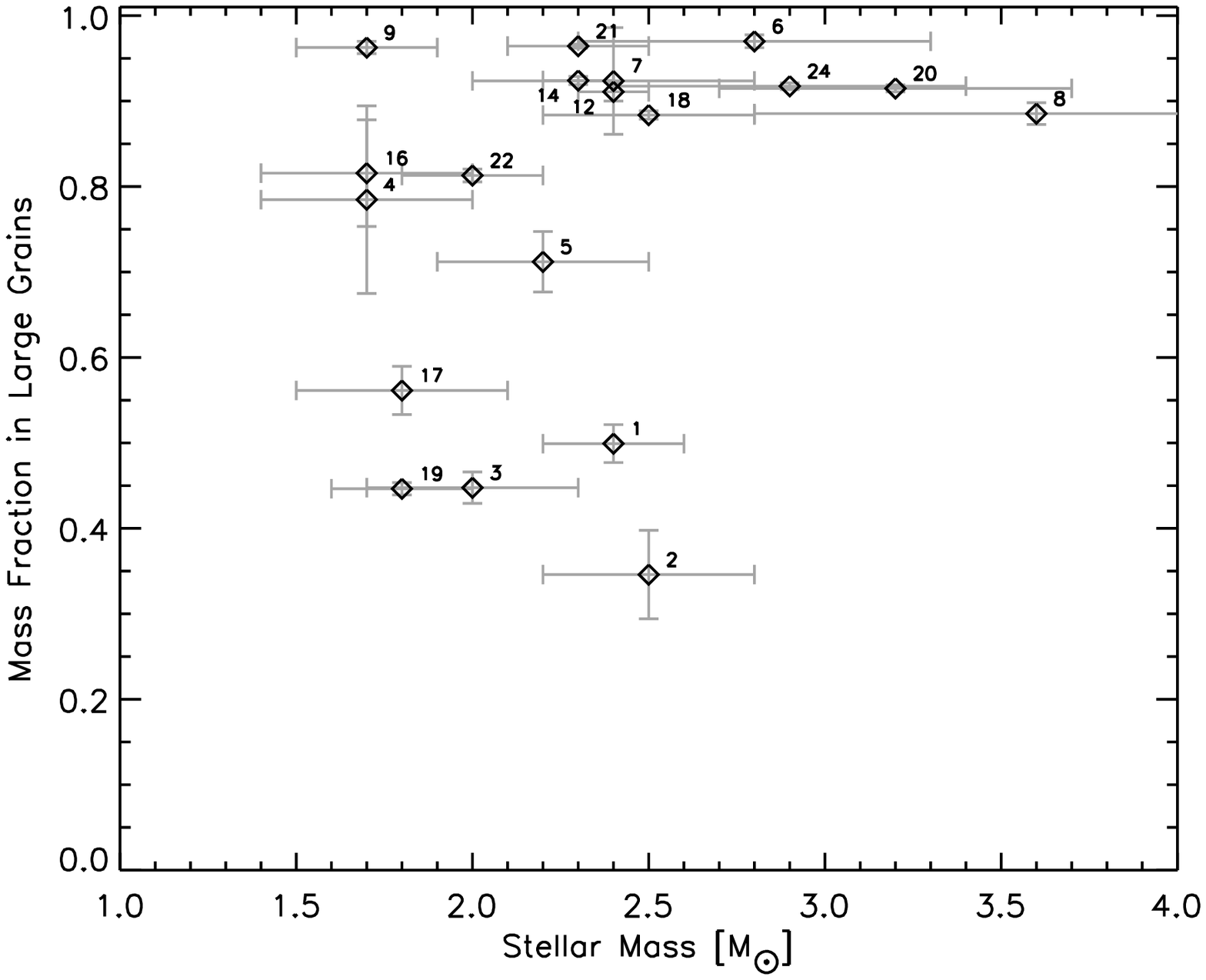}}
  \caption{The mass fraction of dust residing in large grains vs.
the stellar mass. All stars with a mass $M \gtrsim 2.5$\,\msun \ have a mass
fraction of large grains above 85\,\%.}
  \label{fig:t2_fs:Growth_vs_Mstar}
\end{figure}

The fraction of crystalline silicates is correlated with the mass and
luminosity of the central star.  This is visualized in
Fig.~\ref{fig:t2_fs:Xsil_vs_Mstar}. The higher mass (higher
luminosity) stars have an on average higher crystallinity than the
lower mass (lower luminosity) stars. This correlation will be discussed
in section~\ref{sec:t2_fs:new_constraints}.

All stars with a stellar mass above about 2.5\,\msun \ 
($L>$60\,\lsun) have a high fraction ($\gtrsim$85\,\%)
of large grains (Fig.~\ref{fig:t2_fs:Growth_vs_Mstar}).
Possibly, conditions in the disks around more massive stars are more favourable
for growth than they are in the disks around lower mass stars. An alternative
explanation for the observed trend is that
the disks around the more massive stars, which are all relatively
young (Fig.~\ref{fig:t2_fs:Mstar_vs_age}), are more turbulent
than those around the less massive, older stars. 
As a consequence the mixing in the
young disks will be more efficient. While in the older stars the large
grains decouple from the gas and settle to the midplane (and therefore
will not be detected in the 10 micron spectrum anymore), in the young
stars, larger grains may still reach the disk surface and cause the
observed average grain size to be higher.

\section{Discussion}
\label{sec:t2_fs:discussion}

We now discuss our fit results in terms of the processes that are
responsible for the dust evolution.  We will first briefly outline the
expected conditions that prevailed during the active disk phase
(section\,\ref{sec:t2_fs:active_disk_phase}),
i.e. the phase prior to the passive disk phase.  For a comprehensive
review of active disks, see e.g. \cite{2000prpl.conf..377C}.  The disks of
all stars in our sample are in the passive disk phase. The
characteristics of this phase are discussed in
section\,\ref{sec:t2_fs:passive_disk_phase}. In 
section\,\ref{sec:t2_fs:new_constraints} we summarize the constraints
put on dust processing by this work and previous studies. Lastly,
in section\,\ref{sec:t2_fs:global_picture} we sketch a scenario
that is consistent with the current knowledge of dust processing
and disk evolution.

\subsection{The active disk phase}
\label{sec:t2_fs:active_disk_phase}
The accretion of matter onto a forming proto-star is believed to be
initially spherically symmetric. At some point, as the proto-star
contracts, conservation of angular momentum inhibits further spherical
accretion.  The accretion process then proceeds through a disk, and is
accompanied by a bi-polar outflow. The fact that material is radially
transported through the disk implies that the disk has viscosity,
which is in turn coupled to turbulence.

During this so-called \emph{active disk phase}, gravitational energy
of the accreting material is dissipated in the disk, thereby heating
it. Accretion rates in the active phase may reach values up to
10$^{-5}$\,\msunyr \ \citep{2000prpl.conf..377C}.  Close to the central star
the main energy source of the disk is accretion luminosity, whereas
at larger radii irradiation by the star and hot inner disk regions is
expected to be the main heating source of the disk material.

The dust in the disk consists mainly of silicates (studied in this
work), and carbon. The dust does not contribute significantly to the
disk mass in this phase of the disk evolution (the gas over 
dust ratio is on the order of
10$^2$ by mass). However, the thermal radiation emitted by the dust is
the dominant cooling process in the disk. In the outer disk regions
where the main energy source is irradiation, the dust governs
the heating of the disk as well. Therefore,
the dust properties determine the disk temperature, except in the
innermost disk region where the heating is dominated by viscous
dissipation of gravitational energy of the accreting material.
The gas, heated by the dust,
provides the pressure support of the disk.

The densities in the disk are high, especially near the disk
mid-plane, and one may expect coagulation of small dust grains
into larger aggregates to occur. Already after $\sim 10^{4}$ yr of
dynamical disk evolution the average mass of a grain can increase by a
factor $10^{1}$ to $10^{2}$, and close to the disk midplane the grains
may reach sizes of at least a few times 10\,$\mu$m up to millimeters
\citep[][]{2001ApJ...551..461S}. Close to the central star
the temperatures in the disk can reach values in excess of 1\,000\,K. At
this temperature the (initially mostly amorphous) silicates are
annealed, to form crystalline silicates.  Yet closer to the star,
temperatures reach values of about 1\,500\,K, and crystalline silicates
may form by gas phase condensation of evaporated material.

In some of the stars in our sample, the crystalline silicates appear
so prominent that it is unlikely that their emission arises from the
innermost disk region only; the abundance of crystalline silicates
must be high in a relatively large part of the region of the disk responsible for
the 10\,$\mu$m emission (the innermost 10-20\,AU of the disk).
There are in essence two
possible ways to get crystalline material at $\sim$10\,AU distance from the
star: (1) thermal processing in the hot inner disk and subsequent
radial transport of this material outward, and (2) local production of
crystalline material at large distance from the star in transient
heating events caused by shocks \citep[e.g.][]{2002ApJ...565L.109H}
or lightning \citep[e.g.][]{1998A&A...331..121P,2000Icar..143...87D}.
During the active disk phase the
accretion rate in the viscous disk is high and the disk will be
turbulent. This is expected to enable radial mixing of material
\citep[][]{2002A&A...384.1107B,2004A&A...413..571G}.
 
\subsection{The passive disk phase}
\label{sec:t2_fs:passive_disk_phase}
When the supply of fresh material from the maternal cloud has
exhausted, further accretion onto the star proceeds only on a very low
level (order 10$^{-8}$\,\msunyr). Under these conditions the energy
production by viscous dissipation can be fully neglected. Throughout
the disk the temperature of the material is determined by absorption
of stellar radiation; this is referred to as the \emph{passive disk
phase}.  The stars that we study are in this evolutionary
phase.
In the observed SEDs of sources in
the passive disk phase, the infrared excess typically becomes
noticeable above the photospheric emission at a wavelength of about
1\,$\mu$m. The excess emission in the near-infrared indicates that the
inner radius of the disk is determined by the evaporation temperature
of silicates (at about 1\,500\,K). At wavelengths longward of 2\,$\mu$m the
infrared emission from a gas rich disk completely dominates the SED.

The formation of planets and planetesimals is thought to occur during the
passive disk phase. As the disk dissipates on a timescale of
$10^7$\,yr, the infrared excess fades. The inner disk regions become
devoid of gas and dust first.  When the system has evolved into the
\emph{debris disk phase}
\citep[e.g. $\beta$\,Pictoris,][]{1984BAAS...16..483A,1984Sci...226.1421S}
excess emission can be seen above the stellar photospheric emission only 
at wavelengths above $\sim$10\,$\mu$m.

\subsubsection{The era of crystallization}
During the passive disk phase there is a region in the inner disk
where the temperature is above 1\,000\,K, and therefore the silicates
will be crystallized. As there is no significant accretion luminosity,
this region will be smaller than in the active disk phase. It
is unclear whether the disk will be turbulent enough for significant
radial mixing to take place during the passive phase.  Also the
proposed mechanisms for local production of crystalline silicates in
the outer disk regions are more efficient in the active disk phase.
Theoretically therefore, the active disk phase is the preferred era
for the crystallization of the dust.  Nonetheless it has, from an
observational point of view, been suggested that the crystallinity of
HAe disks gradually evolves from low to high values during the passive
phase \citep{2000prpl.conf..613G}.

\subsection{New constraints on dust processing}
\label{sec:t2_fs:new_constraints}

We first briefly repeat the results that follow directly from our spectral
modeling:
\begin{enumerate}

\item All disks have already substantial removal of the smallest
grains (in our fits represented by the 0.1\,$\mu$m grains).
The mass fraction of large (in our fits 1.5\,$\mu$m) grains ranges from
about 35 to almost 100 percent.

\item The crystallinity ranges from about 5 to 30 percent. Disks with a 
high crystallinity are always dominated by large grains.

\item Vice versa, disks with more than 85 percent large grains show a
crystallinity above 10 percent; disks with less than 85 percent large
grains show a crystallinity below 10 percent.

\item Large crystals are found when large amorphous grains are abundant.
We do not find disks in which the bulk of the grains is large, and the
crystalline silicates are small. 

\item In disks with a crystallinity below 15 percent, more than 45 percent
of the crystalline silicates consist of forsterite.
Above 15 percent crystallinity,
more than 45 percent of the crystalline silicates consist of
enstatite.
\end{enumerate}

\noindent
When the stellar parameters of our sample stars 
(see Table\,\ref{tab:t2_fs:source_list}) are taken into account,
several additional conclusions can be drawn:

\begin{enumerate}
\setcounter{enumi}{5}

\item The disks with the highest degree of crystalline material
(between 20 and 30 percent) all belong to stars with a mass above 2.5\,\msun, and a
luminosity above 60\,\lsun.

\item Below a crystallinity of 20 percent, there does not appear to be a
correlation between stellar parameters (mass, luminosity, age) and
crystallinity. Separate studies of the 10 micron spectra of T\,Tauri
stars (Przygodda et al. \citeyear{2003A&A...412L..43P};
Honda, private communication)
indicate that the crystallinities of T\,Tauri star disks are
similar to those observed in the disks of those HAe stars in our sample
that have $M$\,$\lesssim$\,2.5\,\msun \ (i.e. $\lesssim$10 percent). 
We note that a small subset of
the T\,Tauri star disks shows a significantly higher crystallinity,
deviating from the general trend \citep{2003ApJ...585L..59H}.

\item The stars in our sample with masses below 2.5\,\msun \ have ages
ranging from 2$\times$10$^6$ to $\sim$10$^7$ yr (see
Fig.~\ref{fig:t2_fs:Mstar_vs_age}).  If we consider only this
subgroup, we find no relation between the stellar age and the disk
crystallinity.  

\item The stars that lack a 10 micron silicate emission band 
(HD\,97048, HD\,100453, HD\,135344 and HD\,169142) tend to be 
old, with ages of 5-10\,Myr.

\end{enumerate}

\noindent
A number of previous studies have been concerned with the
crystalline silicates in HAe star disks. Here we briefly recall
some of the results that are of relevance for this study:

\begin{enumerate}
\setcounter{enumi}{9}

\item Recent observations using the 10\,$\mu$m instrument MIDI
\citep{2003SPIE.4838..893L} on the VLT
Interferometer \citep{2003SPIE.4838...89G} have enabled the extraction
of the spectra of the innermost $\sim$2\,AU of three HAe stars 
\citep{2004Natur.432..479V}, showing that:
\begin{itemize}
\item[-] HD\,144432 has virtually all its crystals in the inner 2\,AU.
\item[-] HD\,163296 has a flatter crystallinity "gradient" and has
     a higher overall crystallinity than HD\,144432.
\item[-] HD\,142527 has the highest crystallinity of the three
     stars studied, virtually the entire inner 
     disk is crystallized. Also at radii larger than 2\,AU, the abundance
     of crystalline silicates is relatively high.
\item[-] The innermost regions of the disks have experienced more
     grain growth than the outer disk regions. Small grains
     are mostly found at larger distance from these stars.
\end{itemize}

\item ISO observations of HD\,179218 and HD\,100546 show the presence
of cold (100-150\,K) crystalline silicates. This implies a substantial
crystallinity even at distances between 20 and 40-50\,AU
\citep[][ME01]{1998A&A...332L..25M,2001A&A...375..950B}.

\item A 33.5\,$\mu$m forsterite band is tentatively detected in the 
ISO spectrum of HD\,100453 \citep{Vandenbussche2004}.

\end{enumerate}

\vspace{0.3cm}

\noindent

We will now discuss the implications of the points mentioned above.
For clarity, we have labeled each of the implications with the index
number(s) of the point(s) that lead to it.

\begin{description}
\item[1, 2, 3:] The dust in the circumstellar disks coagulates more
easily than it crystallizes.  This is likely due to the fact that the
circumstances enabling grain growth (i.e. high densities) prevail in a
much larger part of the disk than the circumstances needed for
crystallization of silicates (i.e. high temperatures).  Once a certain
level of coagulation is reached, crystallinity becomes a lot more
obvious in the upper layers of the disk.

\item[4:] This observation is consistent with crystallization occurring as
the grains grow: whatever population of grains is
present, is crystallized.
Most of the crystallization occurs when the average grain 
size is already large (micron sized).

\item[1, 2, 3, 10, 12:] Coagulation is more efficient in the inner
disk regions than in the outer disk regions. In the sources that lack
a silicate band coagulation must have proceeded furthest.  Our
relation between grain growth and crystallization
(Fig.~\ref{fig:t2_fs:Xsil_vs_growth}) then implies that these disks
must have a substantial fraction of crystalline silicates, certainly
above 10 percent and perhaps more than 20 percent. However, we do
not find the usual forsterite band at 11.3\,$\mu$m nor the 23.5\,$\mu$m
forsterite band. So, also the crystalline silicates must on average be
large. Indeed this is what was concluded for HD\,100453
\citep{Vandenbussche2004}.  Again, it seems that whatever
(inner disk) grain population is present, is crystallized.

\item[4:] Micron sized crystalline grains are present in the disks.
It is hard to produce crystalline grains of these sizes via shock
heating. At any rate, our data put severe limits on the efficiency of
shock heating, as the shock mechanism must be able to produce
sufficiently strong shocks to crystallize micron sized grains.

\item[5:] If we consider two possible sources of crystalline
silicates, i.e. chemical equilibrium reactions (gas phase condensation
and subsequent gas-solid reactions) and thermal annealing, then we
can note the following \citep{2004A&A...413..571G}:
\begin{itemize}
\item[-] Enstatite is expected to be the dominant crystalline species
formed by chemical equilibrium processes in most of the inner disk,
apart from the hottest, innermost region, which is dominated by forsterite.
\item[-] Since most likely amorphous olivine is the most abundant
species entering the disk from the ISM, forsterite 
(or better, crystalline olivine) is expected to
be the dominant crystalline species formed by thermal annealing.
\end{itemize}
In disks with a low crystallinity, forsterite is the dominant crystalline
species, suggesting that this material is formed by thermal annealing.
The high abundance of enstatite observed in the highly crystalline disks
indicates that the production of crystalline material has occurred by
means of chemical equilibrium processes in these sources.

\item[1, 2, 3, 4, 10:] The TIMMI2 and MIDI data taken together are
consistent with crystallization starting in the innermost
regions. Disks with a high crystallinity in the TIMMI2 data have a
substantial fraction of crystalline grains in the 5-10\,AU area (or
even in the 20-50\,AU region, see e.g. the ISO spectra of HD\,100546, HD\,179218
\citep{1998A&A...332L..25M,2001A&A...375..950B}),
i.e. well outside the regions where thermal annealing and chemical
equilibrium processes are expected to be effective.

\item[5, 11:] HD\,179218 has a high abundance of cold enstatite. In
the context of the above reasoning, this enstatite must have been
produced either locally in transient heating events, or transported
outward from the inner disk to observable cold disk regions.  
If chemical equilibrium processes
are an important means of producing enstatite, the HD\,179218 data
point to enstatite production in the inner regions and transport to
larger distance by radial mixing.

\item[6:] Crystallization is more efficient in disks around higher
mass, higher luminosity stars. Several possible explanations for this
trend can be identified:
\begin{enumerate}
\item during the active phase, the region in which the disk material
is sufficiently hot for thermal annealing (and chemical equilibrium
processes) to occur is larger in high mass stars, since the energy
dissipated in the disk per unit mass accreted material is higher. If
this is the dominant effect, our results favour crystallization in
the active phase.
\item during the passive phase, when the extent of the region where
annealing (and chemical equilibrium processes) can occur is set by
irradiation from the central star, the higher luminosity of the more
massive stars causes this region to be larger than in the low mass
stars. If this is the dominant effect, our results favour
crystallization in the passive phase.
\item the disks around more massive, more luminous stars may be more
turbulent, both during the active and passive phase. The outward
mixing of processed material will then be more efficient in the
high mass stars, causing a more prominent appearance of this dust
in the spectrum.
\end{enumerate}
It is currently unclear which of the above effects is dominant.
The study of young ($\tau < 10^6$\,yr), low mass stars may allow
to answer this question.

\vspace{0.15cm}

\item[8:] This observation suggests that whatever caused the observed range
of crystallinities in stars with a mass below about 2.5\,\msun, 
occurred before 2$\times$10$^6$
yr. This puts the epoch of crystallization in the active and/or early
passive phase.
\end{description}

\vspace{0.3cm}

\noindent
\subsection{Global picture that emerges}
\label{sec:t2_fs:global_picture}
The 10\,$\mu$m emission is emitted by the surface layer of the disk
at radii below 20\,AU from the central star. In the midplane of
this region (and presumably even at larger radii), dust coagulation
takes place. Vertical mixing ensures that the coagulated, micron-sized
grains reach the disk surface layer and become apparent in the spectrum.
Growth is most efficient in the innermost disk regions
where the densities are highest.
Crystallization by means of thermal annealing and chemical equilibrium
processes occurs in the hot innermost disk region. In many disks
there are significant amounts of crystalline material at larger
($\gtrsim$5\,AU) distances from the star, where thermal annealing and
chemical equilibrium processes are ineffective. The combined evidence
presented in section~\ref{sec:t2_fs:new_constraints} seems to favour
radial transport from the innermost disk regions as the source of
these crystals, above local production mechanisms (e.g. shock
annealing).

Crystallization is most efficient in the disks surrounding the more
massive, more luminous stars (M\,$\gtrsim 2.5$\,\msun,
L\,$\gtrsim$\,60\,\lsun).  In these disks, the region in which thermal
annealing and chemical equilibrium processes can produce crystalline
silicates is larger than in the disks surrounding lower mass stars.
Additionally, the disks around the more massive stars may be more
turbulent, enabling more efficient radial mixing.  It is unclear
whether the crystallization of the dust in the disks occurs
predominantly in the active disk phase, or in the passive phase that
follows. Our data do suggest that the disks reach their final
crystallinity relatively early in the passive phase 
($\tau$\,$\lesssim$\,2\,Myr). Therefore, crystallization happens during the active
and/or early passive phase.

\section{Conclusions}
\label{sec:t2_fs:conclusions}
We have undertaken a large spectroscopic survey of Herbig Ae stars in the
10\,$\mu$m atmospheric window, and have presented new spectra of 23 stars.
The correlation between the shape and the strength of the 10\,$\mu$m silicate 
emission band reported by \cite{2003A&A...400L..21V}
is reconfirmed with a larger sample. We have performed compositional fits
to the silicate feature using opacities of minerals commonly found in
circumstellar disks (amorphous olivine and pyroxene, crystalline olivine
and pyroxene, and amorphous silica) and polycyclic aromatic hydrocarbons. 
For all minerals we have allowed both ``small'' (0.1\,$\mu$m) and ``large''
(1.5\,$\mu$m) grains in order to study the effects of grain growth.
This set of materials is sufficient to reproduce the wide variety of
observed spectral shapes.

We find a trend between the mass fraction in large grains and the mass
fraction in crystalline grains: \emph{all sources with a high
crystallinity have a high mass fraction in large grains}.  There are
no highly crystalline sources which are dominated by small grains.
Most sources have a mass fraction in large grains of more than 80
percent.

We note that there is an important bias in our sample (and probably in
most samples of Herbig stars studied in the literature): the more
massive sample stars ($\sim$3\,\msun, ``high mass'') are younger than
the less massive ($\lesssim$\,2.5\,\msun, ``low mass'') stars.
There are no low mass stars younger than 1\,Myr in our sample.  In
order to establish the disk conditions at the end of the active disk
phase, which precedes the passive phase, it is essential that
such very young, low mass stars are found and studied. Since these
stars may still be enshrouded in circumstellar material, they may
have to be selected using infrared data.

We find a trend between the derived crystallinity of the dust and the
mass (and luminosity) of the central star: the disks around stars with
a mass larger than $2.5$\,\msun \ (a luminosity above 60\,\lsun) have
a higher crystallinity ($\gtrsim$\,20\,\%) than the less massive, less
luminous stars.  Within the subset of sources with a stellar mass
below $2.5$\,\msun, no correlation between crystallinity and stellar
parameters (mass, luminosity, age) is seen. These lower mass stars in
our sample are all older than $\sim2$\,Myr. Since in this subset there
is no correlation between age and crystallinity, we conclude that the
crystallization of the material predominantly happens in the active or
early passive disk phase (before $2$\,Myr).

The evidence presented in this paper combined with conclusions from
other studies seems to favour a scenario in which crystalline
silicates are produced in the innermost regions of the disk and
transported outwards. Spatially resolved spectra of these disks, as
can be obtained using for example the MIDI instrument on the Very 
Large Telescope Interferometer, will provide crucial information on the radial
dependence of the mineralogy of the dust in these disks. In addition,
measurements at longer wavelengths can probe colder regions (further
out) in the disk which can provide further constraints on the
temperature structure and spatial distribution of the dust.

\begin{acknowledgements}
This research has made use of the SIMBAD database,
operated at CDS, Strasbourg, France. We gratefully acknowledge
J.\,W.~Hovenier for thorough reading and 
valuable comments on an earlier version of the manuscript.
\end{acknowledgements}

\bibliographystyle{aa}
\bibliography{references}

\end{document}